\documentclass[11pt]{article}  
\usepackage{jheppub} 
\usepackage{epsfig}
\usepackage{color}
\usepackage{amsmath}
\usepackage{amssymb}
\usepackage{latexsym}
\usepackage{dsfont}
\usepackage{verbatim}
\usepackage{bm}% bold math
\usepackage{multirow}
\usepackage[T1]{fontenc}

\newcommand{\be}{\begin{equation}}
\newcommand{\ee}{\end{equation}}
\newcommand{\bea}{\begin{eqnarray}}
\newcommand{\eea}{\end{eqnarray}} 
\newcommand{\bel}{\begin{align}}
\newcommand{\eel}{\end{align}}
 
\newcommand{\nn}{\nonumber}
\newcommand{\nnn}{\nonumber \\}
\newcommand{\nm}{\nonumber \\ &&}
\newcommand{\nnnza}{\nonumber \\
	&& \qquad \qquad}
\newcommand{\nnnzad}{\nonumber \\
	&& \qquad \qquad  \qquad}

\newcommand{\Gep}{\textmd{G}_{1}^{+}}
\newcommand{\Gem}{\textmd{G}_{1}^{-}}
\newcommand{\Gdp}{\textmd{G}_{2}^{+}}
\newcommand{\Gdm}{\textmd{G}_{2}^{-}}
\newcommand{\Hp}{\textmd{H}^{+}}
\newcommand{\Hm}{\textmd{H}^{-}}
\newcommand{\Aep}{\textmd{A}_{1}^{+}}

\newcommand{\Adp}{\textmd{A}_{2}^{+}}
\newcommand{\Adm}{\textmd{A}_{2}^{-}}
\newcommand{\Tep}{\textmd{T}_{1}^{+}}
\newcommand{\Tem}{\textmd{T}_{1}^{-}}
\newcommand{\Tdp}{\textmd{T}_{2}^{+}}
\newcommand{\Tdm}{\textmd{T}_{2}^{-}}
\newcommand{\Ep}{\textmd{E}^{+}}
\newcommand{\Em}{\textmd{E}^{-}}

\begin{document}
 
\title{ Lattice operators for scattering of  particles with spin}

\author[a,b,c]{S. Prelovsek,}
\author[b]{U. Skerbis} 
\author[d]{and C.B. Lang}
\affiliation[a]{Faculty of Mathematics and Physics,  University of Ljubljana, Jadranska 19, 1000 Ljubljana, Slovenia}
\affiliation[b]{Jozef Stefan Institute, Jamova 39, 1000 Ljubljana, Slovenia}
\affiliation[c]{Theory Center, Jefferson Lab,12000 Jefferson Avenue, Newport News, VA 23606, USA}
\affiliation[d]{Institute of Physics, University of Graz, Universit\"atsplatz 3, A--8010 Graz, Austria}

\emailAdd{sasa.prelovsek@ijs.si}
\emailAdd{ursa.skerbis@ijs.si}
\emailAdd{christian.lang@uni-graz.at}

\abstract{  We construct operators for simulating the scattering of two hadrons with spin on the lattice. Three methods are shown to give the consistent operators for $PN$, $PV$, $VN$ and $NN$ scattering, where $P$, $V$ and $N$ denote pseudoscalar, vector and nucleon. Explicit expressions for operators are given for all irreducible representations at lowest two relative momenta.  
 Each hadron has a good helicity in the first method. The hadrons are in a certain partial wave $L$ with total spin $S$ in the second method. These enable the physics interpretations of the operators  obtained from the general projection method. The correct transformation properties of the operators in all three methods are proven.   The total momentum of two hadrons is restricted to zero since parity is a good   quantum number in this case.   }

\keywords{ lattice field theory, two-hadron scattering, interpolating fields } 

\arxivnumber{1607.06738 }
%\pacs{11.15.Ha, 12.38.Gc} 

\maketitle 
 
\section{Introduction}

The study of two-hadron interactions with lattice QCD requires operators that create and annihilate the two-hadron system with the desired quantum numbers.  Systems of two hadrons with zero spin have been extensively studied and the corresponding scattering matrices  for interactions of two pseudoscalars ($P$) were extracted, for example $\pi\pi$, $K\pi$, .... The simulations of two-hadron systems where one or both hadrons carry non-zero spin are scarce, and those mostly focused on the partial-wave $L=0$. An example of a study that aimed also at higher partial waves considered nucleon-nucleon scattering \cite{Berkowitz:2015eaa}. 

There is a great need for lattice results on further channels or higher partial waves where one or both hadrons carry spin. We have in mind  channels that involve the  nucleon $N$ or/and vectors $V=J/\psi,~\Upsilon_b,~D^*,B^*,...$ which are (almost) stable under strong interactions. The $PN$ or $VN$ scattering is crucial for ab-initio study of baryon resonances and pentaquark candidates,  $PV$ is essential for mesonic resonances and tetraquark candidates, while $NN$  to grasp two-nucleon interaction and deuterium\footnote{ The operators below can be straightforwardly generalized also to  channels which involve scattering of particles with $J^P=0^+,~1^+, ~\tfrac{1}{2}^-$.  Most of these hadrons decay strongly in Nature. }. We consider the most common  simulations that are done in a box of size $L$ with periodic boundary conditions in space. The momenta $p$ of non-interacting single hadrons are multiples of $2\pi/L$ in this case.

 We focus on the system with total momentum zero where parity is a good quantum number. This has a practical advantage since fixing the parity gives a strong handle on which channels and partial waves can contribute as eigenstates in a given study.  On the lattice the continuum symmetry is reduced  to a discrete symmetry and one extracts the eigenstates that transform according to the chosen irreducible representation (irrep) $\Gamma$ of the discrete group $O^{(2)}_h$.  Only the states with desired parity contribute to  a given irrep when the total momentum is zero, while both even and odd partial waves contribute to a given irrep when the total momentum is not zero.

In this paper we  present the general expressions to construct operators for scattering of  particles with spin using three methods. The corresponding proofs of the correct transformation properties are given. 
Then the explicit expressions  $H^{(1)}(p)H^{(2)}(-p)$ for $PV$, $PN$, $VN$ and $NN$ are presented for lowest two relative momenta $|p|=0,~2\pi/L$. 

 The reason to introduce two-hadron operators in a lattice simulation is to extract  the eigenstates and their energies in a desired scattering channel. These energies  render the scattering   phase shift via the well-known L\"uscher method \cite{Luscher:1990ux}, which was originally derived for the two spin-less particles. The relation has been generalized to the scattering of two particles with arbitrary spin 
 \cite{Briceno:2014oea,Luu:2011ep,Gockeler:2012yj,Bernard:2012bi,Briceno:2013lba} and the most general case is considered in \cite{Briceno:2014oea}.   
This can be directly applied to extract the scattering matrix from  the energies obtained using the operators presented in this work.  

 The three methods to construct operators complement each other and it is verified that all three methods lead to consistent results.   The projection method is a general mathematical tool which leads to one or several operators $O_{\Gamma,r,n}$ that transform according to given irrep $\Gamma$ and row $r$. It is a ``bottom-up'' method, working entirely with lattice operators. But it gives little guidance on which continuum quantum numbers of total  $J$, spin $S$, orbital  momentum $L$ or single-particle helicities $\lambda_{1,2}$ will be related with a given operator. This is remedied with the helicity and partial-wave ``top-down'' methods where first the operators with good continuum quantum numbers  $(J,P,\lambda_{1,2})$ or $(J,L,S)$ are constructed and then subduced to the irreps $\Gamma$ of the discrete group.   The hope is that the results indicate which linear combinations $O_{\Gamma,r,n}$ of various $n$ enhance couplings to the states with desired continuum quantum numbers. An analogous strategy for single-hadron states was proposed by the Hadron Spectrum Collaboration in \cite{Dudek:2010wm}. 
We do not prove whether  two particle operators also retain a ``memory'' of their construction origin. Future simulations   analogous  to the single-hadron case \cite{Dudek:2010wm} will be necessary to explore how well that may be realized in practice. 

 Certain aspects of constructing the lattice operators for scattering of particles with  spin were already presented before: partial-wave method in  \cite{Berkowitz:2015eaa,Wallace:2015pxa}, projection method   for example in \cite{Gockeler:2012yj},  helicity states for single-hadrons in  \cite{Thomas:2011rh} and some considerations for two hadrons in \cite{Dudek:2012gj,Wallace:2015pxa}. 
The study \cite{Moore:2006ng} combines single particle irreps to two-particle operators. For this one needs the values of all corresponding Clebsch-Gordan coefficients to embed the product irreps of the single particle
operators in the irreps of the two-particle frame (see Sect. \ref{sec:single_particle_irreps}). Despite all previous work, various practical aspects and proofs are lacking  to build a two-hadron operator   related  to desired continuum quantum numbers in a given irrep. Here we devise  two-hadron helicity operators, verify their consistency  with other two methods and provide related proofs. We also argue how simple non-canonical single-hadron operators can be used as building blocks for that purpose.

 % \vspace{0.1cm}
   
The paper is organized  as follows. The single-hadron operators that are employed to build two-hadron operators are considered in Section \ref{sec:single_hadron}. The required  transformations properties of two-hadron  operators   are given in Section \ref{sec:two_hadron}. The general expressions for two-hadron operators    with three methods are presented in Section \ref{sec:three_methods}, while the proofs of correct transformation properties are given in Appendix \ref{sec:proofs}. The explicit  operators for lowest two relative momenta are collected in Sections \ref{sec:results} and \ref{sec:HH}. The necessary technicalities are  delegated to Appendix \ref{sec:technicalities}.
 
 \section{Single-hadron operators and their transformations}\label{sec:single_hadron}
 
 The single-hadron operators $H(p)$ need to have certain transformation properties under rotations $R$ and inversion $I$ in order to build two-hadron operators   $H^{(1)}(p)H^{(2)}(-p)$ with desired transformation properties.  
  The  states and creation operators  $H^\dagger_{m_s}(p)$  with spin $s$ transform as 
  \begin{align}  \label{1}
   &R|p,s,m_s\rangle=\sum_{m_s'} D_{m_s'm_s}^s(R)|Rp,s,m_s'\rangle\;, \qquad  I |p,s,m_s\rangle=(-1)^P |-p,s,m_s\rangle\\
   &RH^\dagger_{m_s}(p)R^{-1}=\sum_{m_s'} D_{m_s'm_s}^s(R) H_{m_s'}^\dagger (Rp)\;,\qquad  I H^\dagger_{m_s}(p)I = (-1)^P H^\dagger_{m_s}(-p)\,.\nonumber
 \end{align}
 For a particle at rest $m_s$ is a good quantum number of the spin-component $S_z$.  
 The $m_s$ is generally not a good quantum number for $H_{m_s}(p\not = 0)$; in this case it denotes the  eigenvalue of $S_z$ for the corresponding field $H_{m_s}(0)$, which has good $m_s$.
 The transformations of annihilation operators $H_{m_s}(p)$ are obtained by the hermitian conjugation   
  \be \label{2}
RH_{m_s}(p)R^{-1}= \sum_{m_s'} D_{m_sm_s'}^s(R^{-1}) H_{m_s'} (Rp)\;,\qquad  I H_{m_s}(p)I = (-1)^P H_{m_s}(-p)\;,~
  \ee
  where the following properties of the Wigner D-matrix  and R are used throughout this work 
  \be\label{3}
  D(R)=D^\dagger(R^\dagger)\;,\qquad R^\dagger=R^{-1}\;, \qquad D(R_1R_2)=D(R_1)D(R_2)\;.
  \ee 
  We employ  the conventional definition of $D$ used  in \cite{Jacob:1959at,pdg14} and discussed in Appendix \ref{sec:wignerD}. 
  Note that the $z$-component of spin $m_s$  is a good quantum number for a particle at rest.   
  
The pseudoscalar fields $P$ ($J^P=0^-$), vector fields $V$ ($J^P=1^-$) and nucleon fields $N$ ($J^P=\tfrac{1}{2}^+$) are considered, together with their mutual two-hadron scattering  channels later on.  One can choose the simplest annihilation operators that transform according to (\ref{2}), for example 
            \begin{align}
    \label{6}
   & P(p)=\sum_x \bar q(x) \gamma_5 q(x) e^{ipx}\\
  &  V_{m_s=\pm 1}(p)=\frac{\mp V_x(p)+iV_y(p)}{\sqrt{2}},\  \ V_{m_s=0}(p)=V_z(p),\quad V_i(p)=\sum_x \bar q(x) \gamma_i q(x) e^{ipx}, \ i=x,y,z\nonumber\\
  & N_{m_s=1/2}(p)= {\cal N}_{\mu=1}(p)\;, \ {\cal N}_{m_s=-1/2}(p)= {\cal N}_{\mu=2}(p)\;,   \ \  {\cal N}_\mu(p)\!=\!\sum_x \epsilon_{abc} [q^{aT}(x) C \gamma_5 q^b (x)] ~q^c_{\mu}(x)~e^{ipx},   \nonumber
    \end{align} 
        where $p$ and $x$ are three-vectors and the time index is omitted.  The ${\cal N}_{1,2}$ are the upper two components of Dirac four-spinor ${\cal N}_{\mu=1,..,4}$ in the Dirac basis. The relations for the  vector states $|V_{m_s= 1}\rangle=(- |V_x\rangle-i|V_y\rangle)/\sqrt{2}$ versus annihilation operators $V_{m_s= 1}=(- V_x+iV_y)/\sqrt{2}$ are discussed in Appendix \ref{sec:vectors}. The operators $P$, $N_{1/2,-1/2}$ and $V_{i=x,y,z}$ like  (\ref{6}) will be basic building blocks of our two-hadron operators.
  
  Since all  two-hadron operators will entail transformations of annihilation fields $H=P,V,N$, we present them explicitly
    \begin{align}
     \label{7}
    & RP(p)R^{-1}=P(Rp),\quad IP(p)I=-P(-p)\\
  &R V_i(p)R^{-1}=  T^{s=1}_{ji}(R)^* V_{j}(Rp)=\exp(-i \vec n \vec J \omega)_{ji}V_{j}(Rp),\quad IV_i(p)I=-V_i(-p)\quad  i,j=x,y,z\nonumber \\
  & R N_{m_s}(p) R^{-1}= D_{m_s'm_s}^s(R)^* N_{m_s'}(Rp)=[\exp(-\tfrac{i}{2} \vec n \vec \sigma  \omega)]^*_{m_s'm_s} N_{m_s'}(Rp) ,\quad  I N_{m_s}(p) I=N_{m_s}(-p)  \nonumber
  \end{align}
  with $m_{s,s'}=\pm 1/2$  and $(J_k)_{ij}=-i\epsilon_{ijk}$ rendering real $T^{s=1}$. The fields transform as basis-vectors, which is contrasted to the transformation of components in Appendix \ref{sec:components}. The transformations of $V_i(p)$ in (\ref{7}) can be used to verify the transformations (\ref{2}) of the fields $V_{  m_s}(p)$ defined in (\ref{6}).  
  
Note that   canonical hadron fields $H_{m_s}^{(c)}(p)\equiv L(p) H_{m_s}(0)$ also transform under rotations as given in (\ref{2})\footnote{See for example Eq. (2.5.23) of Weinberg \cite{weinberg:qft} with $W({\cal R},p)={\cal R}$. }, but those would be less practical to implement. The canonical fields are obtained from the rest fields $H_{m_s}(0)$   (which have good $m_s$) with the boost $L(p)$ from $0$ to $p$. The boost of $V_{m_s=1}(0)$ to $p_x\propto e_x$ would  render $V_{m_s=1}^{(c)}(p_x)=[-\gamma V_x(p_x)+iV_y(p_x)]/\sqrt{2}$ which depends on $\gamma=(1-v^2)^{1/2}$. The boost  of $N_{1/2}(0)={\cal N}_1(0)$ to $p_x$ would render   $N_{1/2}^{(c)}(p_x)\propto {\cal N}_1(0)+\tfrac{p_x}{m+E}{\cal N}_4$, which contains also lower components and depends on the mass $m$.
We will employ the simpler   fields $V$ and $N$ (\ref{6}) that have all the required transformation  properties  (\ref{2},\ref{7}) to prove the correct transformations of the  two-hadron operators in Appendix \ref{sec:proofs}. The fields (\ref{6}) agree with the canonical fields for $p=0$.
    
 \section{Transformation properties of two-hadron operators}\label{sec:two_hadron}
 
 We consider only two-hadron states with total momentum zero $P_{tot}=0$ that have good parity $P$. 
 These also have good total spin $J$ and its $z$-component $m_J$ in continuum, where the annihilation operators have to  transform as (\ref{2})
 \be
 \label{4}
   RO^{J,m_J}(P_{tot}\!=\!0)R^{-1}= \sum_{m_J'} D_{m_Jm_J'}^J(R^{-1}) O^{J,m_J'} (0)\,\;  R\in O^{(2)}, \;\;  I O^{J,m_J}(0)I = (-1)^P O^{J,m_J}(0)~.
  \ee
  Such continuum-like operators will present an intermediate step below and the only relevant 
  rotations will be those contained in the discrete group. 
  
 On the cubic lattice the continuum rotation group is reduced  to the cubic group $R\in O$ with 24 elements for integer $J$.
These are given in terms of the rotation angle $-\pi\leq \omega <\pi$ around vector $\vec n$ in positive direction and we use the ordering $i=1,..,24$  listed in Table A.1 of \cite{Bernard:2008ax}. The double-cover group $O^2$ has the same 24 elements, and additional 24 elements ($i=25,..48$)\footnote{Our ordering and range of $\omega$ for $O^2$ differs from Table A.3 in  \cite{Bernard:2008ax}.}  with the same $\vec n$ and angle $\pi\leq \omega <3\pi$ where $\omega_{i+24}=\omega_i+2\pi$.  The number of symmetry elements gets doubled to $\tilde R=\{R,IR\}\in O^{(2)}_h$ when  inversion is a group element. 

The representations (\ref{4}) with given $J$ and $m_J$ are reducible under $O^{(2)}_{h}$, so we seek the 
       states/operators that transform according to the corresponding irreducible representation (irrep)  $\Gamma$ and row $r$
  \begin{align}
  \label{5}
  &R|\Gamma,r\rangle = \sum_{r'} T^{\Gamma}_{r',r}(R)|\Gamma,r'\rangle \quad  R\in O^{(2)},\qquad I|\Gamma,r\rangle =(-1)^P |\Gamma,r\rangle\;,\nonumber \\ 
 &RO_{\Gamma,r}R^{-1} = \sum_{r'} T^{\Gamma}_{r,r'}(R^{-1})O_{\Gamma,r'} \quad  R\in O^{(2)}  ,\qquad IO_{\Gamma,r}I =(-1)^P O_{\Gamma,r}\;.
     \end{align}
     The systems with integer $J$ transform according to irreps $\Gamma=A_{1,2}^\pm,~E^\pm,~T_{1,2}^\pm$, while systems with half-integer $J$ according to   $\Gamma=G_{1,2}^\pm~,H^\pm$ as summarized in Table \ref{tab:irreps}. 
The upper index stands for parity and the number of rows $r=1,..,$dim$_{\Gamma}$ is equal to the dimensionality of irrep $\Gamma$.  We employ the same conventions for rows in all irreps as in \cite{Bernard:2008ax}\footnote{Those are partly discussed in our Appendix \ref{sec:rows}}. The explicit representations $T^{\Gamma}_{r,r'}(R)$ for $R\in O^{(2)}$ are given in Appendix A of \cite{Bernard:2008ax}\footnote{In \cite{Bernard:2008ax} the $T^{\Gamma}_{r',r}$   are denoted by $(R_i)_{\alpha\beta}$. }.
          
     \begin{table}[h!]
 	\begin{center}
 		\begin{tabular}{c|c}
 			J &  $\Gamma$ (dim$_\Gamma$)  \\ 
 			\hline
 			$0$  & $A_{1}(1)$  \\ 
 			$\frac{1}{2}$ &  $G_{1}(2)$ \\ 
 			$1$ & $T_{1}(3)$ \\ 
 			$\frac{3}{2}$ &  $H(4)$ \\ 
 			$2$ &  $E (2) \oplus T_{2}(3)$ \\ 
	 		$\frac{5}{2}$ &  $ H(4) \oplus G_{2}(2)$ \\ 
 			$3$&  $A_{2}(1) \oplus T_{1} (3)\oplus T_{2}(3)$
 		\end{tabular} 
 	\end{center}
 	 	\caption{ Continuum spins $J$, corresponding lattice irreps $\Gamma$ and their dimension dim$_\Gamma$ for group $O^{(2)}$. \label{tab:irreps}
}
 \end{table}     
   
 As indicated above, here we consider only two-hadron states with total momentum zero. 
    Both even and odd partial waves contribute to a given irrep when total momentum is not zero. Working with operators that mix parity is no problem in principle and has already been considered for, e.g., pseudoscalar-pseudoscalar scattering. It leads to more eigenstates in the given irreducible representation and one has to apply further assumptions and tools to extract the scattering phase shift.  Note that in both cases, exact or mixed parity, the lattice irreps may still couple to a tower of higher partial waves. The phase shifts for $\ell\leq 2$ in the parity mixed case have been successfully extracted, for example, via the parameterization of the S-matrix  by the Hadron Spectrum Collaboration in \cite{Wilson:2014cna,Moir:2016srx}.         
  
  \section{Two-hadron operators in three methods}\label{sec:three_methods}
  
Here we present  two-hadron operators with total momentum zero that transform according to (\ref{5}). They will be constructed by three methods, where continuum-like operators (\ref{4}) will appear as an intermediate  step in two of the methods.  Their  correct transformation properties (\ref{4},\ref{5}) are proven in Appendix \ref{sec:proofs}. 
      
  \subsection{Projection method}
  
 This is a well-known mathematical method, where a projector to the desired irrep $\Gamma$ and row $r$ is used on an arbitrary operator (see for example Section 4.19 of  \cite{Dawber:symmetries}) 
  \begin{align}
  \label{O_P}
 & ||p|,\Gamma,r,n\rangle=\sum_{\tilde R\in O^{(2)}_h} T^\Gamma _{r,r}(\tilde R)^*~\tilde R ~[|2,-p,s_2\rangle^{a}|1,p,s_1\rangle^{a}] \nonumber\\
 & O_{|p|,\Gamma,r,n}=\sum_{\tilde R\in O^{(2)}_h} T^\Gamma _{r,r}(\tilde R)~\tilde RH^{(1),a}(p)H^{(2),a}(-p)\tilde R^{-1}~,\qquad   n=1,..,n_{max}\;.
  \end{align}
 
  The $H^{a}$  are  arbitrary single hadron lattice operators of desired $|p|$ and any  $p$ and $m_s$, for example operators (\ref{6}) or their linear combinations; the other index $(1)$ and $(2)$ denotes the particle type.  We have taken $H^a$ with  all possible combinations of  direction $p$ and polarizations of both particles $m_{s1}$ and $m_{s2}$ (for vectors we chose  $V_x$, $V_y$ and $V_z$ as $H^a$).  For fixed $|p|$, $\Gamma$ and $r$ one can get one or more linearly independent operators  $O_{|p|,\Gamma,r,n}$ which are indicated by $n$. 
 As an illustration we present    two-linearly independent $PV$ operators    
  $$O_{|p|=1,T_1^+,r=3,n=1}\propto \!\!\!\sum_{p=\pm e_z}\!\! \text{P}(p) V_z(-p) \;,\ \      O_{|p|=1,T_1^+,r=3,n=2}\propto \!\!\!\!\!\!\sum_{p=\pm e_x,\pm e_y} \!\!\!\!\!\!\! \text{P}(p) V_z(-p)\;,\quad n_{max}=2\;,$$
while others  are listed in the Section \ref{sec:results} and Appendix \ref{sec:HH}. 

 The $\tilde R\in O^{(2)}_h $ in (\ref{O_P}) is the operator symbolising rotations $R$ possibly combined with inversions ($R$ is reserved in this article for rotations only) and one uses in practice  
  \be
 \tilde RH^{(1)}H^{(2)}\tilde R^{-1}= \tilde RH^{(1)}\tilde R^{-1} ~\tilde RH^{(2)}\tilde R^{-1},\qquad \tilde R\in O^{(2)}_h
 \ee   
 where the action of rotation $R$ or inversion $I$ on all $H=P,V,N$ are given in (\ref{2},\,\ref{7}).
   The representation matrices  $T^{\Gamma}(\tilde R)$ for all elements $\tilde R$ are listed  for all irreps in  Appendix A of \cite{Bernard:2008ax}.  The proof of correct transformation is shown for completeness in Appendix \ref{sec:proofs}.

The projection method  is very general, but it does not offer  physics intuition what 
$O_{|p|,\Gamma,r,n}$ with different $n$ represent in terms of the continuum quantum numbers. This will be remedied with the next two methods, that indicate which linear combinations of $O_{n}$ correspond in the continuum to certain partial waves or helicity quantum numbers.

  \subsection{Helicity method}

The states with good helicity in the continuum scattering have been thoroughly discussed by Jacob and Wick \cite{Jacob:1959at}. They have been considered for one-hadron states on the lattice by the Hadron-Spectrum collaboration \cite{Thomas:2011rh}, but they have not been widely used for two-hadron scattering on the lattice yet. 

The $m_s$ is a good quantum number for particles with $p=0$ or $p_z\propto e_z$ since $L_z=0$ and $S_z=J_z$, while it is generally not  a good quantum number for arbitrary $p\not =0$. Here $L_z$, $S_z$ and $J_z$ denote components of the orbital, spin and total angular momentum, respectively.  The advantage of  the helicity $h$
\be
\label{h}
 h\equiv \frac{S\cdotp p}{|p|}
 \ee
 is that it is a good quantum number for arbitrary $p$, where $S$ and $p$ are three-vectors.   To obtain a helicity state, one starts from a state with momentum  $p_z\propto e_z$ and $|p_z|=|p|$ that has good $m_s$. This state is   rotated from $p_z$ to desired direction of  $p$ with $R_0^p$,
 \be
 \label{13}
|p,s,\lambda\rangle^{h}\equiv R_0^p|p_z,s,m_s\rangle\,,\qquad H_{\lambda}^{h}(p)\equiv  R_0^p ~H_{m_s=\lambda}(p_z) ~(R_0^p)^{-1},\qquad p_z\propto e_z,\quad |p_z|=p~.
\ee  
 The upper index $h$ indicates that  the polarisation index  ($\lambda$) stands for the helicity of the particle (not $m_s)$. Different choices of $R_0^p$ are possible, which lead to different phases in the definition of $|p,s,\lambda\rangle^{h}$. We do not restrict to a specific choice of $R_0^p$ since our resulting two-particle operator will be independent of this (except for the overall phase of operator, which is irrelevant).  Simple examples of $H_{p_z,m_s=\lambda}$ can be read-off from (\ref{6}) and the action of $R_0^p$ on them is given by (\ref{7}). 

The arbitrary rotation $R$ rotates the momentum $p$ and the spin $S$ in the same way,  so the helicity $h\propto S\cdotp p$ of the state remains the same and  only the direction of the momentum changes \cite{Jacob:1959at,Thomas:2011rh}
\be
\label{12}
R|p,s,\lambda\rangle^{h}= e^{-i\varphi(R)} |Rp,s,\lambda\rangle^{h},  \qquad  RH_{\lambda}^{h}(p) R^{-1}= e^{i\varphi(R)} H_{\lambda}^{h}(Rp)~,
\ee 
where only the phase $\varphi(R)$ depends on $R$.

\subsubsection{Construction of two-hadron helicity operators}

The starting point is a two-particle state composed of  $|p,s_1,\lambda_1\rangle^{h}$ and  $|\!-\!p,s_2,\lambda_2\rangle^{h}$, where $p$ can be chosen as arbitrary momentum in a given shell $|p|$, and $\lambda_{1,2}$ are chosen helicities. The two-particle helicity state and the annihilation operator with the correct transformation properties under rotation (\ref{4}) are
\begin{eqnarray}
\label{O_helicity_noP}
||p|,J,m_J,\lambda_1,\lambda_2,\lambda\rangle^{h}&=& \sum_{R\in O^{(2)}} D^J_{\lambda,m_J}(R^{-1}) R\,[|p,s_1,\lambda_1\rangle^{h}| \!-\!p,s_2,\lambda_2\rangle^{h}]\nonumber\\
O^{|p|,J,m_J,\lambda_1,\lambda_2,\lambda}&=& \sum_{R\in O^{(2)}} D^J_{m_J,\lambda}(R) ~RH_{\lambda_1}^{(1),h}(p) H_{\lambda_2}^{(2),h}(-p)R^{-1}
\end{eqnarray}
with the proof for operators given in Appendix \ref{sec:proofs}.   Note that the labels on the left-hand side should just indicate the origin of construction; they would correspond to the continuum quantum numbers if $R$ was summed over the continuum rotation group.
The main difference with  respect to the continuum case 
\cite{Jacob:1959at} is the discrete sum over the elements of the discrete group $R\in O$ for integer $J$ and  $R\in O^{2}$ for half-integer $J$. Appendix   \ref{sec:wignerD} provides $D$-matrices \footnote{One has  to use the same rotation  in $D$ and in the transformation of fields. The additional  rotation of $2\pi$ that is present in half of the $O^2$ elements  renders  factors $(-1)^J$ from $D^J(R)$ and $(-1)^{s_1}(-1)^{s_2}$ from the transformations of the both fields in  (\ref{O_helicity_noP}).  }  for $R$ in $O$ and double-cover $O^2$. The helicity of the two-particle state is $\lambda=\lambda_1-\lambda_2$ for all cases we consider\footnote{In the continuum this follows from the continuum integration over Euler angle $\gamma$ \cite{Jacob:1959at}.  }.  

Note that the operator (\ref{O_helicity_noP}) is a sum of states with fixed   $\lambda_{1,2}$
since the helicity of each particle is conserved with rotation (\ref{12}), 
\be
O^{|p|,J,m_J,\lambda_1,\lambda_2,\lambda}= \sum_{R\in O^{(2)}} D^J_{m_J,\lambda}(R) ~e^{i\Phi(R)}~H_{\lambda_1}^{(1),h}(Rp) H_{\lambda_2}^{(2),h}(-Rp)~.
\ee
The phase $\Phi(R)$ depends also on the choice of $R_0$ through the definition of the helicity state (\ref{13}). We do not explicitly determine $\Phi(R)$, but rather directly employ expressions (\ref{O_helicity_noP}) and (\ref{O_helicity}) to get explicit results for the interpolators in Section \ref{sec:results}. 

The final    helicity operator with desired parity $P=\pm 1$ is obtained from (\ref{O_helicity_noP}) by parity projection $\tfrac{1}{2}({\cal O}+P I {\cal O} I)$    
\begin{align}
\label{O_helicity} 
O^{|p|,J,m_J,P,\lambda_1,\lambda_2,\lambda}= \frac{1}{2}\!\!\sum_{R\in O^{(2)}} D^J_{m_J,\lambda}(R) ~&RR_0^p~
[H^{(1)}_{m_{s_1}=\lambda_1}(p_z) H^{(2)}_{m_{s_2}=-\lambda_2}(-p_z)\\
&\ +PIH^{(1)}_{m_{s_1}=\lambda_1}(p_z) H^{(2)}_{m_{s_2}=-\lambda_2}(-p_z)I]~(R_0^p)^{-1}R^{-1}~,\nonumber
\end{align} 
where we have expressed $H^{h}$ (\ref{13}) with fields $H_{m_s}(p_z)$ (\ref{6}) that have good  $m_s$. The actions of inversion  $I$ and the rotation $R$ on the fields $H_{m_s}$ are given in (\ref{7}). One chooses particular $p$ with given $|p|$ and performs rotation $R_0^p$ from $p_z$ to  $p$. There are several possible choices of $R_0^p$, but they lead only to different overall phases for the whole operator (\ref{O_helicity}), which is irrelevant\footnote{The $p$ is the same for all terms  (\ref{O_helicity}),  so the phase related to the choices of $R_0^p$ is the same in all terms.}.  

 The   operators for $|p|=1$ (in units of $2\pi/L$) will be explicitly presented in Section \ref{sec:results}. We take the simplest choice $p=p_z=(0,0,1)$ and    $R_0^p=\mathbf{1}$  in (\ref{O_helicity})\footnote{Such a choice is not available for some shells, for example $|p|^2=2$. In this case one chooses some $p$ available in the shell and evaluates (\ref{O_helicity}).  } 
\begin{align}
\label{O_helicity_p1} 
O^{|p|=1,J,m_J,P,\lambda_1,\lambda_2,\lambda}\!\!=  \frac{1}{2}\!\!\sum_{R\in O^{(2)}} D^J_{m_J,\lambda}&(R) ~R[H^{(1)}_{m_{s1}=\lambda_1}(p_z) H^{(2)}_{m_{s_2}=-\lambda_2}(-p_z)\\
& \ +PIH^{(1)}_{m_{s1}=\lambda_1}(p_z) H^{(2)}_{m_{s_2}=-\lambda_2}(-p_z)I]~R^{-1}~.\nonumber
\end{align} 
As an illustration we present two $PV$ operators with the same $J^P$ ,
$$O^{|p|=1,J=1,m_J=0,P=+,\lambda_V=0}\propto \!\!\!\sum_{p=\pm e_z} \text{P}(p) V_z(-p) ~,\ \       O^{|p|=1,J=1,m_J=0,P=+,\lambda_V=1}\propto \!\!\!\!\sum_{p=\pm e_x,\pm e_y} \!\!\!\!\text{P}(p) V_z(-p)~. $$

\subsubsection{ Subduction  to irreducible representations }\label{sec:subduction}

The   operators (\ref{O_helicity}) would correspond to the irreducible representations only for the continuum rotation group $R\in SO(3)$ in which  case $J$ indicates the continuum quantum number. These operators (\ref{O_helicity}) represent reducible representation under the reduced discrete group $O^{(2)}$. In simulations  one needs to  employ operators, which transform according to irreducible representations $\Gamma$ and row $r$ of $G=O^{(2)}$. Those are obtained from the continuum-like operators $O^{J,m_J}$ by the subduction  \cite{Dudek:2010wm,Edwards:2011jj} 
  \be
  \label{14}
  O_{|p|,\Gamma,r}^{[J,P,\lambda_1,\lambda_2,\lambda]}=\sum_{m_J}  {\cal S}^{J,m_J}_{\Gamma,r} O^{|p|,J,m_J,P,\lambda_1,\lambda_2,\lambda}\;.
       \ee 
The subduction coefficients ${\cal S}$ are real and are given in   Appendices
     of \cite{Dudek:2010wm} and \cite{Edwards:2011jj} for all irreps.   For the $T_{1,2}$ irreps our convention for rows is different and the corresponding coefficients ${\cal S}$ are listed in Appendix \ref{sec:rows}. 
          
     This strategy, to start from operators with continuum spin $O^{J,m_J}$ and  subduce them to irreps $O^{[J]}_{\Gamma,r}$, was proposed by the Hadron Spectrum Collaboration in \cite{Dudek:2010wm}.
      It was then extensively employed  for single hadrons by the same collaboration\footnote{This collaboration employed operators for two spin-less particles, which were obtained by combining $O_{\Gamma_1,r_1}\times O_{\Gamma_2,r_2}\to O_{\Gamma,r}$ using corresponding Clebsch-Gordan coefficients  \cite{Dudek:2012gj}. }. The subduced operators $O^{[J]}_{\Gamma,r}$ were found to ``carry the memory'' of the spin $J$, where they were subduced from and dominantly couple to the lattice eigenstates with this spin.   Whether that expectation holds for two-particle operators combining spin and angular momentum is unclear; if the
 behaviour is like for the single-particle operators one
may expect that the subduced operators  $O_{|p|,\Gamma,r}^{[J,P,\lambda_1,\lambda_2,\lambda]}$ will  dominantly represent eigenstates with  continuum quantum numbers $J,\lambda_1,\lambda_2,\lambda$.

  \subsection{Partial-wave method}  
 
Often one is interested in the scattering of two hadrons in a given partial wave $L$. 
   The orbital angular momentum $L$ and total spin $S$ are not separately conserved, so several $(L,S)$ combinations can render the same $J$, $m_J$ and $P$, which are good quantum numbers.   Nevertheless, the $L$ and $S$ are valuable physics quantities to label continuum annihilation field operator   
\be
\label{O_LS}
O^{|p|,J,m_J,L,S}=\sum_{m_L,m_S,m_{s1},m_{s2}} C^{Jm_J}_{Lm_L,Sm_S}C^{Sm_S}_{s_1m_{s1},s_2m_{s2}} \sum_{R\in O} Y^*_{Lm_L}(\widehat{Rp}) H^{(1)}_{m_{s1}}(Rp)H^{(2)}_{m_{s2}}(-Rp)~.
\ee
  The operator has parity $P=P_1P_2(-1)^L$ and its correct transformation property under rotation (\ref{4}) is demonstrated in Appendix \ref{sec:proofs}.   The operator was considered for nucleon-nucleon scattering already in \cite{Berkowitz:2015eaa}, where the proof was not presented (this reference uses $Y_{Lm_L}$ where we have $Y_{Lm_L}^*$).  The $C$ are Clebsch-Gordan coefficients, $p$ is an arbitrary momentum with desired $|p|$, $Y_{Lm_L}(\widehat{Rp})$ is a spherical harmonic for direction $(\theta,\phi)$ given by the  normalized vector $\widehat{Rp}$. Simple choices of one-particle annihilation operators $H$ are listed in (\ref{6}).    The sum is over $R\in O$ for integer as well as half-integer $J$, since rotations for angle $\omega$ and  $\omega+2\pi$ have the same effect on $p$. Here is an example of two $PV$ operators with the same $J$ and $S$ 
  \begin{align}
  O^{|p|=1,J=1,m_J=0,L=0,S=1}&\propto \!\!\!\!\!\!\sum_{p=\pm  e_x,\pm e_y, \pm e_z}\!\!\!\!\!\!\! \text{P}(p) V_z(-p) \;,\nonumber\\
     O^{|p|=1,J=1,m_J=0, L=2,S=1}&\propto \!\!\! \sum_{p=\pm  e_x,\pm e_y } \!\!\!\!\!\!\text{P}(p) V_z(-p) - 2 \!\! \sum_{p=\pm e_z } \!\text{P}(p) V_z(-p)\;. \nonumber
  \end{align}

The  operators (\ref{O_LS}) that are labeled by the continuum $J$  are reducible under  reduced  group $O^{(2)}$, as discussed in section \ref{sec:subduction}. The  operators that transform according to irrep $\Gamma$ and row $r$ of $G=O^{(2)}$ are obtained from $O^{J,m_J}$ by the subduction  \cite{Dudek:2010wm}
  \be
  \label{15}
  O_{|p|,\Gamma,r}^{[J,L,S]}=\sum_{m_J}  {\cal S}^{J,m_J}_{\Gamma,r} O^{|p|,J,m_J,L,S}\;,
       \ee 
       where ${\cal S}$ are the same subduction coefficients as in (\ref{14}).       
One expects that the subduced operators  $O_{|p|,\Gamma,r}^{[J,L,S]}$ carry the memory of 
      continuum $J,L,S$ and  dominantly couple to eigenstates with these  quantum numbers. This depends on how close one is to the physical limit and whether there may be an actual physical transition\footnote{For example $L=0$ and $L=2$ mix in continuum in the $PV$ and $NN$ channel with  $S=1$ and $J^P=1^+$.} between
states of different $(L,S)$ in the given chanel $J^P$.   If this transition is small, one can hope that operators (\ref{15}) would be valuable  handles for   certain lower values of $J$ and $L$, which are of most practical importance. 
      Certain higher $L^\prime>L$ inevitably lead to the same operators (or their linear combinations), and $O_{\Gamma}^{[J,L,S]}$ could provide an effective handle on the low partial wave $L$ particularly if the corresponding higher partial waves  $L^\prime$ are negligible (as  illustrated in Section \ref{sec:higherL}).        
      
\subsection{Relation between partial-wave and helicity operators}
      
Operators calculated in partial-wave method (\ref{O_LS}) are  linear combinations of operators calculated by helicity method (\ref{O_helicity})
      \begin{equation}
\label{O_rel}
 O^{|p|,J,m_J,S,L}  = \sqrt{\frac{2L+1}{4 \pi}} \sum_{\lambda = -S}^{S} \sum_{\lambda_{1},\lambda_{2}} \sum_{\lambda^{'}} D_{\lambda^{'},\lambda }^{J}(R_0^p)  C^{J\lambda}_{L0,S \lambda} C_{s_1\lambda_1,s_2\;-\lambda_2}^{S\lambda}~ O^{|p|,J,m_J,\lambda^{'},\lambda_1,\lambda_2} 
 \end{equation}
where the proof of the relation  is presented in the Appendix \ref{sec:relation}. For  $|p|=1$, we take the reference momentum $p=R_0^pp_z$   along the z-direction, so $R_0^p$  and  $D^{J}(R_0^p)$ are identity matrices.
            
 \subsection{Method combining single particle irreducible representations}\label{sec:single_particle_irreps}
      
Let us mention also  a fourth method, which is not applied here. This approach   \cite{Moore:2006ng} combines single particle irreps $\Gamma_{1,2}$ (for the cubic group at $p=0$ or for the little groups at $p\not = 0$)   to two-particle operators in irrep $\Gamma$: $\Gamma_1\otimes \Gamma_2 \to \Gamma$. One hopes that this way one may profit from the optimized single particle operators for given $\Gamma_{1,2}$, as for example in the $PP$ study  \cite{Wilson:2014cna}.  One needs the values of all corresponding Clebsch-Gordan coefficients to embed the product irreps of the single particle operators in the irreps of the two-particle frame. The method has been, for example, used in $PP$ scattering \cite{Wilson:2014cna} where the Clebsch-Gordan coefficients are known \cite{Dudek:2012gj}.  The reference \cite{Moore:2006ng} provides which Clebsch-Gordan coefficients are non-zero for scattering of particles with spin.  
This approach  does not relate the resulting operators to continuum quantum numbers like $L$ or  $S$.   Nevertheless, it  might turn out to be  particularly valuable for constructing two-hadron operators with non-zero total momenta.

  \section{Operators with momenta $|p|=0,1$}\label{sec:results}
  
  The explicit expressions for annihilation operators $H^{(1)}(p)H^{(2)}(-p)$ for $PV$, $PN$, $VN$ and $NN$ scattering in three methods are collected here and in Appendix \ref{sec:HH}. We restrict  to lowest two momenta $|p|=0,1$ (in units of $2\pi/L$) for the individual hadrons on the lattice of size $L$ with periodic boundary conditions in space.  
  Operators from the projection method $O_{|p|,\Gamma,r,n}$  will be given first, where $n=1,..,n_{max}$ indicate all  linearly independent combinations for given $\Gamma$ and $r$. 
  Then operators from helicity method $O_{|p|,\Gamma,r}^{[J,P,\lambda_1,\lambda_2,\lambda]}$ and partial-wave method $O_{|p|,\Gamma,r}^{[J,L,S]}$ will be expressed as  linear combinations of known operators $O_{|p|,\Gamma,r,n}$ obtained by the projection method. This indicates which linear combinations have to be employed in a simulation to enhance a coupling to the state with continuum quantum numbers $J,P,\lambda_1,\lambda_2$ or $J,L,S$. This coupling   is expected to be favourable when the  studied system is approaching the physical limit, if certain higher partial waves are suppressed and if the physical mixing of different $(L,S)$ in given $J^P$ channel is small. 
  
   In both, helicity and partial wave, methods we consider $J\leq 2$ for integer and  $J\leq 5/2$ for half-integer $J$. 
The  helicity operator will not be given for $|p|=0$  as helicity $h$ (\ref{h}) is intended for non-zero momentum.  This operator satisfies 
 \be
 \label{16}
 O_{\Gamma,r}^{[J,P, \lambda_1 ,\lambda_2,\lambda=\lambda_1-\lambda_2]}\propto O_{\Gamma,r}^{[J,P, -\lambda_1 ,-\lambda_2,-\lambda]}
 \ee 
 since $h$ (\ref{h}) reverses sign if $p\to -p$ and $S\to S$, while $p\to -p$ only multiplies the operator by $\pm 1$ depending on parity.     The $m_J$ will appear as an index in helicity and partial-wave operators in cases when only one $m_J$ is present in subduction  to given irrep and row (\ref{14},\ref{15}). The vector fields are expressed in the $V_{x,y,z}$ basis (\ref{6}), which are most straightforward for implementation. The normalization of each operator is arbitrary. For brevity, only the first row $r=1$ is displayed, which should suffice for an actual simulation. Other rows and $|p|$ can be obtained from the general expressions for $O$ above. 
 
We verified that  all partial-wave and helicity operators are related by (\ref{O_rel}).  
Also, we confirmed that the number of linearly  independent operators in each irrep agrees with  the fourth method  \cite{Moore:2006ng} in Section \ref{sec:single_particle_irreps}, as elaborated for the $PV$ case  below. 
 The number of operators based in this method can be deduced based on  Table I (for $p=0$) and Table III (for $|p|=1$) of   \cite{Moore:2006ng}.
    
Before listing the operators, let us point out again the key expressions that have been used to derive them.   The operators from projection method are obtained using (\ref{O_P}) with representations $T^{\Gamma}(\tilde R)$ listed  in  Appendix A of \cite{Bernard:2008ax}. The  helicity operator (\ref{O_helicity_p1}) for $|p|=1$ is  subduced to irreps by (\ref{14}), while the partial-wave operator (\ref{O_LS}) comes from the subduction with (\ref{15}). The simple examples of the fields $H_{m_s}$ are collected in (\ref{6}) with their transformations in (\ref{2},\ref{7}) and  Wigner-D functions in (\ref{wignerD}).

  \subsection{$PV$ operators}
\subsubsection{$|p|=0$}
 \bea
 && T_1^+:\qquad\qquad\qquad\qquad\qquad\qquad\qquad\qquad\qquad\qquad\qquad\qquad\qquad\qquad\qquad\qquad ~ \\
 &&O_{T_1^+,r=1}= \text{P}(0) V_x(0)
 \nonumber\\
 && O_{T_1^+,r=1}^{[J=1,L=0,S=1 ]}= O_{T_1^+,r=1} \nonumber
 \eea
Operators for other irreps are equal to $0$. 

\subsubsection{$|p|=1$}\label{sec:higherL}

  \bea
 &&
 A_1^-:\label{A1m}\qquad\qquad\qquad\qquad\qquad\qquad\qquad\qquad\qquad\qquad\qquad\qquad\qquad\qquad\qquad\qquad\\
&& O_{A_1^-,r=1}= \text{P}(e_x) V_x(-e_x)-\text{P}(-e_x) V_x(e_x)+\text{P}(e_y)  V_y(-e_y) -\text{P}(-e_y) V_y(e_y) \nnnza
 + \text{P}(e_z) V_z(-e_z)-\text{P}(-e_z) V_z(e_z) \nonumber \\
&& O_{A_1^-,r=1}^{[J=0,m_{J}=0 ,P=-, \lambda_V=0,\lambda_P=0]}=O_{A_1^-,r=1}^{[J=0,m_{J}=0,L=1,S=1 ]}= O_{A_1^-,r=1}  \nn
\eea
\bea
&&
 T_1^+: \qquad\qquad\qquad\qquad\qquad\qquad\qquad\qquad\qquad\qquad\qquad\qquad\qquad\qquad\qquad\qquad\label{T1p}\\
&& O_{T_1^+,r=1,n=1}=\text{P}(e_x) V_x(-e_x)+\text{P}(-e_x) V_x(e_x)  \nonumber \\
&& O_{T_1^+,r=1,n=2}=\text{P}(e_y) V_x(-e_y)+\text{P}(-e_y) V_x(e_y)+\text{P}(e_z) V_x(-e_z)+\text{P}(-e_z) V_x(e_z)  \nonumber \\ 
&& O_{T_1^+,r=1}^{[J=1,P=+, \lambda_{V}= \pm 1,\lambda_P=0]}=  O_{T_1^+,r=1,n=2}    \nonumber\\
&& O_{T_1^+,r=1}^{[J=1,P=+, \lambda_{V}=0 ,\lambda_P=0]}=  O_{T_1^+,r=1,n=1}      \nonumber \\
&&O_{T_1^+,r=1}^{[J=1,L=0,S=1 ]}= O_{T_1^+,r=1,n=1} + O_{T_1^+,r=1,n=2}   \nonumber \\
&& O_{T_1^+,r=1}^{[J=1,L=2, S=1] }= -2 ~ O_{T_1^+,r=1,n=1} + O_{T_1^+,r=1,n=2}   \nn
\eea
\bea
&&
 T_1^-: \qquad\qquad\qquad\qquad\qquad\qquad\qquad\qquad\qquad\qquad\qquad\qquad\qquad\qquad\qquad\qquad\\
&& O_{T_1^-,r=1}=-\text{P}(e_y) V_z(-e_y)+\text{P}(-e_y) V_z(e_y)+\text{P}(e_z) V_y(-e_z)-\text{P}(-e_z) V_y(e_z) \nonumber \\
&& O_{T_1^-,r=1}^{[J=1,P=-, \lambda_{V}= \pm 1,\lambda_P=0]}=O_{T_1^-,r=1}^{[J=1,L=1,S=1] }= O_{T_1^-,r=1} \nn
\eea
\bea
&&
 T_2^+:\qquad\qquad\qquad\qquad\qquad\qquad\qquad\qquad\qquad\qquad\qquad\qquad\qquad\qquad\qquad\qquad \\
&& O_{T_2^+,r=1}=\text{P}(e_y) V_x(-e_y)+\text{P}(-e_y) V_x(e_y)-\text{P}(e_z) V_x(-e_z)-\text{P}(-e_z) V_x(e_z) \nonumber \\
&& O_{T_2^+,r=1}^{[J=2,P=+, \lambda_{V}= \pm 1,\lambda_P=0]}=O_{T_2^+,r=1}^{[J=2,L=2,S=1]}= O_{T_2^+,r=1}  \nn
\eea
\bea
&&
 T_2^-: \label{T2m} \qquad\qquad\qquad\qquad\qquad\qquad\qquad\qquad\qquad\qquad\qquad\qquad\qquad\qquad\qquad\qquad\\
&& 
O_{T_2^-,r=1}=\text{P}(e_y) V_z(-e_y)-\text{P}(-e_y) V_z(e_y)+\text{P}(e_z) V_y(-e_z)-\text{P}(-e_z) V_y(e_z) \nonumber \\
&& O_{T_2^-,r=1}^{[J=2,P=-, \lambda_{V}= \pm 1,\lambda_P=0]}=O_{T_2^-,r=1}^{[J=2,L=1,S=1]}=O_{T_2^-,r=1}^{[J=2,L=3,S=1]}=O_{T_2^-,r=1}  \nn
\eea
\bea
&&
E^-: \qquad\qquad\qquad\qquad\qquad\qquad\qquad\qquad\qquad\qquad\qquad\qquad\qquad\qquad\qquad\qquad\\
&& O_{E^-,r=1}=\text{P}(e_x) V_x(-e_x)-\text{P}(-e_x) V_x(e_x)+\text{P}(e_y) V_y(-e_y)-\text{P}(-e_y) V_y(e_y) \nonumber \\
&& \qquad \qquad - 2 \text{P}(e_z) V_z(-e_z)+2 \text{P}(-e_z) V_z(e_z) \nonumber \\
&& O_{E^-,r=1}^{[J=2,P=-, \lambda_{V}= 0,\lambda_P=0]}=O_{E^-,r=1}^{[J=2,L=1,S=1]}= O_{E^-,r=1}^{[J=2,L=3,S=1]}=O_{E^-,r=1} \nonumber\\
&&\nonumber\\
&&  O_{A_1^+}= O_{A_2^+}= O_{A_2^-}= O_{E^+}=0  \;.
\eea 

There is only one independent operator with $|p|\leq 1$ for each irrep (and given row), except for $T_1^+$. The result confirms, for example,  that    $O_{A_1^-,r=1}^{[J=0,P=-, \lambda_V=0]}=O_{A_1^-,r=1}$ (\ref{A1m}) relates to a  channel with negative parity and zero-helicity vector. Or that    $O_{T_2^-,r=1}^{[J=2,L=1,S=1]}=O_{T_2^-,r=1}$ (\ref{T2m}) gives information on   the channel  with $J=2$, $L=1$ and $S=1$.  Of course, certain  states with higher $J$ and $L$ inevitably contribute to the same discrete irrep along the lines of Table \ref{tab:irreps}. For example states with $J=2$, $L=3$, $S=1$ also contribute to the same irrep $T_2^-$ and the operators 
$O_{T_2^-,r=1}^{[J=2,L=1,S=1]}=O_{T_2^-,r=1}^{[J=2,L=3,S=1]}$ can not distinguish between $L=1$ and $L=3$.

The irrep $T_1^+$  presents   an interesting case where there are two independent interpolators  $n=1,2$ at $|p|=1$. The result (\ref{T1p})  indicates that $O_{T_1^+,n=1}$ is  relevant for $\lambda_V=0$, while  $O_{T_1^+,n=2}$ is relevant for  $|\lambda_V|=1$.   The expressions (\ref{T1p}) also indicate which linear combinations of $O_{T_1^+,n}$  need to be employed to study $L=0$ or $L=2$ partial waves. Note that both partial waves contribute to the same $J^P=1^+$ channel even in the continuum $PV$ scattering with $S=1$. Specific higher $J\geq 3$ and $L\geq 4$ lead to the same $T_1^+$ operators (or their linear combinations). In the limit when partial waves $L\geq 4$ are negligible, we expect that $O_{T_1^+}^{[J=1,L=0,S=1 ]}$ and $O_{T_1^+}^{[J=1,L=2,S=1 ]}$  are valuable handles on lowest partial waves $L=0$ and $2$, respectively.    This is expected to work better when the physical transition $(L=0,S=1)\leftrightarrow (L=2,S=1)$ in this channel is small, which  is realized, for example, in the heavy-light meson  observables due to the heavy quark symmetry\footnote{One  $Q\bar q$ state  with $J^P=1^+$ decays only via $L=0$ in the $m_Q\to \infty$ limit \cite{Isgur:1991wq},  and  $O_{T_1^+}^{[J=1,L=0,S=1 ]}$ was used to consider $D_1(2430)\to D^*\pi$ in \cite{Mohler:2012na}. The other $J^P=1^+$ state  decays only via $L=2$  and is narrow.}.    

The correct number of independent interpolators for $|p| = 1$ can be verified as follows.    The total number can be obtained by counting  different $P(p)V_i(-p)$, which span a basis with $6\cdot 3=18$ terms ($6$ for directions of $p$ and $3$ for polarization of $V$). This agrees with the number  $5\cdot 3+2+1=18$ of  interpolators above in all rows  ($5$ for number of independent $O_{T_{1,2},r=1}$ and $3$ for number of rows in $T_{1,2}$). The number of $P(1)V(-1)$ interpolators for each irrep  also agrees with the fourth method  \cite{Moore:2006ng}, discussed in Section \ref{sec:single_particle_irreps}.  The Table I  in \cite{Moore:2006ng}  indicates that $P(1)$ is in irrep $A_2$ of group $Dic_4$, while $V(-1)$ can be in 
  $E_2$ or $A_1$ of  $Dic_4$. The $P(1)V(-1)$ are combined to $A_2\otimes (E_2\oplus A_1)=(T_1^+\oplus T_1^-\oplus T_2^+\oplus T_2^-)\oplus (T_1^+\oplus A_1^-\oplus E^-)$ according to Table III in  \cite{Moore:2006ng}. This agrees with our interpolators above, so $T_1^+$ indeed appears twice.  
  
\subsection{$PN$, $VN$ and $NN$ operators} 

The explicit expressions for $|p|=0,1$ and all irreps are presented in Appendix \ref{sec:HH}. Several linearly independent operators $O_{\Gamma,r,n}$ typically arise for $VN$ and $NN$ in most of irreps. The number of linearly independent operators  in a given irrep is the same as the number of different $(L,S)$ combinations   that can  contribute to given $J^P$ in the continuum limit (this applies for $J\leq 3/2$ which are contained in  a single irrep according to Table \ref{tab:irreps}).
 The partial-wave and helicity operators  indicate which linear combinations of $O_{\Gamma,n}$ are most relevant to study scattering with given $(J,L,S)$ or given $(J,P,\lambda_1,\lambda_2)$. 
 
\section{Conclusions}

We consider  three  different methods to construct  two-hadron interpolators where one or both hadrons carry spin. The focus is on the case with total-momentum zero where parity $P$ is a good quantum number. The correct transformation properties are proven for all three methods. The projection method is a general mathematical tool which leads to one or several operators $O_{\Gamma,r,n}$  that transform according to given irrep $\Gamma$ and row $r$, but it does not give much insight on the underlying continuum quantum numbers.  The partial-wave  and the helicity methods indicate which linear combinations $O_{\Gamma,r,n}$ of various $n$ have to be employed  in order to hopefully enhance couplings to the states with desired continuum quantum numbers.  
The partial-wave method renders operators $O_{\Gamma,r}^{[J,L,S]}$ with enhanced couplings to two-hadron states in partial wave $L$, total spin $S$ and total angular momentum $J$. The helicity method provides operators $O_{\Gamma,r}^{[J,P,\lambda_1,\lambda_2]}$  where each hadron has good helicity $\lambda_{1,2}$. The quality of these enhancements depends on how close one is to the physical limit, the smallness of the disregarded
higher partial waves, and on the magnitude of the  physical transitions between various $(L,S)$ for a given $J^P$. 

Explicit expressions for $PV$, $PN$, $VN$ and $NN$  operators with lowest two relative momenta are provided in all irreducible representations.   We verified  that all three methods lead to consistent results, where partial-wave and helicity methods are particularly helpful for physics interpretation.  The operators shall be valuable to simulate the  scattering of particles with spin in desired channels using quantum field theory on the lattice.
     
     \vspace{1cm}
     
     {\bf Acknowledgments}
We want to thank kindly  R. Brice\~no, J. Dudek, R. Edwards, A. Nicholson, M. Padmanath and A. Walker-Loud  for valuable discussions. We are especially grateful to M. Padmanath for insightful discussions on the construction of  $PN$ interpolators. This work is supported in part by the  Slovenian Research Agency ARRS, and by the Austrian Science Found project FWF:I1313-N27. S.P. acknowledges support from U.S. Department of Energy Contract No. 
DE-AC05-06OR23177, under which Jefferson Science Associates, LLC, manages and operates Jefferson Laboratory.

\appendix
  
  \section{Proofs of  transformation properties for operators }\label{sec:proofs}
  
   \subsection{Projection method}
   
   The arbitrary operator $H^{(1),a}H^{(2),a}$ in (\ref{O_P}) is a linear combination of $O_{\Gamma',r'}$ and the projector projects out $O_{\Gamma,r}$:  
  \bea
\sum_{\tilde R\in G} T_{r,r}^\Gamma(\tilde R) ~\tilde RO_{\Gamma',r'}\tilde R^{-1}
&=& \sum_{\tilde R} T_{r,r}^\Gamma(\tilde R) ~\sum_{r''} T^{\Gamma'}_{r'',r'}(\tilde R)^*O_{\Gamma',r''}\nonumber\\
&=&\sum_{r''} \bigl[ \sum_{\tilde R} T_{r,r}^\Gamma(\tilde R) T^{\Gamma'}_{r'',r'}(\tilde R)^*\bigr] O_{\Gamma',r''}\nonumber\\
 &=&\frac{\mathrm{dim}_\Gamma}{n_G} \sum_{r''} \delta_{\Gamma\Gamma'} \delta_{rr''} \delta_{rr'} O_{\Gamma',r''}= \delta_{\Gamma\Gamma'}  \delta_{rr'} O_{\Gamma,r}\;,
 \eea
 where (\ref{5}) and   the orthogonality theorem \cite{Dawber:symmetries} were used, while $n_G$ is number of group elements.

  \subsection{Helicity method}
  
 Let us  verify that annihilation operator  (\ref{O_helicity_noP}) transforms under rotation $R_a\in O^{(2)}$  as   (\ref{4}), 
  where indices $|p|$ and $\lambda$ are omitted for brevity 
  \bea
R_a O^{J,m_J,\lambda_1,\lambda_2}R_a^{-1} 
&=&  \sum_{R\in O^{(2)}} D^J_{m_J,\lambda}(R) 
R_aR \,H_{\lambda_1}^{h}(p)H_{ \lambda_2}^{h}(-p)R^{-1}R_a^{-1}\\
&=& 
\sum_{R\in O^{(2)}} D^J_{m_J,\lambda}(R_a^{-1}R^\prime) ~
 R^\prime  H_{\lambda_1}^{h}(p)H_{ \lambda_2}^{h}(-p) R^{\prime -1}\nonumber\\
&=&  \sum_{R^\prime\in O^{(2)}} \sum_{m_J^\prime }D^J_{m_J,m_J^\prime}(R_a^{-1})D^{J}_{m_J^\prime,\lambda}(R^\prime) ~
 R^\prime  H_{\lambda_1}^{h}(p)H_{ \lambda_2}^{h}(-p) R^{\prime -1}\nonumber\\
&=& \sum_{m_J^\prime}D^J_{m_J,m_J^\prime}(R_a^{-1})~\sum_{R^\prime\in O^{(2)}} D^{J}_{m_J^\prime,\lambda}(R^\prime) ~
 R^\prime  H_{\lambda_1}^{h}(p)H_{ \lambda_2}^{h}(-p) R^{\prime -1}\nonumber\\
 &=& \sum_{m_J^\prime}D^J_{m_J,m_J^\prime}(R_a^{-1})~O^{J,m_J^\prime,\lambda_1,\lambda_2}\;,\nonumber
\eea 
where we defined $R'=R_aR$ and used (\ref{3}).

  \subsection{Partial-wave method}

  Here we verify that operator  (\ref{O_LS}) transforms under rotation $R_a\in O^{(2)}$  as   (\ref{4}).  First the summation $R\in O$  (\ref{O_LS}) can be replaced by $R\in O^{(2)}$ since both have the same effect on $p$ \footnote{An irrelevant overall factor $1/2$ that might arise is here omitted. }  
\bea
\label{10}
R_aO^{J,m_J,L,S}R_a^{-1}&=&\hspace*{-12pt}\sum_{m_L,m_S,m_{s1},m_{s2}} \!\!\!\!\!\!\!\! C^{Jm_J}_{Lm_L,Sm_S}C^{Sm_S}_{s_1m_{s1},s_2m_{s2}}  \sum_{R\in O^{(2)}} Y^*_{Lm_L}(\hat{Rp})~R_aH_{m_{s1}}(Rp) H_{m_{s2}}(-Rp)R_a^{-1}\nonumber\\
&=&\hspace*{-12pt}\sum_{m_L,m_S,m_{s1},m_{s2}} C^{Jm_J}_{Lm_L,Sm_S}C^{Sm_S}_{s_1m_{s1},s_2m_{s2}}  \sum_{R\in O_h} Y^*_{Lm_L}(\hat{Rp})\nonumber\\
&& \times \sum_{m_{s1}'}D_{m_{s1}m_{s1}'}^{s_1}(R_a^{-1}) H_{m_{s1}'}(R_aRp)\sum_{m_{s2}'}D_{m_{s2}m_{s2}'}^{s_2}(R_a^{-1})H_{m_{s2}'}(-R_aRp)\;,
\eea
where (\ref{2}) was used. 
We intruduce introduce $R'\equiv R_aR$, which is also an element of $O^{(2)}$, and use  
\be
Y^*_{Lm_L}(Rp)=Y^*_{Lm_L}(R_a^{-1}(R'p))=\sum_{m_L'} D^L_{m_Lm_L'}(R_a^{-1})Y^*_{Lm_L'}(R'p)\;,
     \ee  
 where the second step follows from  $Y^*_{Lm_L}(R_1p)=\sum_{m_L'} D^L_{m_Lm_L'}(R_1)Y^*_{Lm_L'}(p)$ \footnote{See for example Appendix C of [Messiah]. This relation depends on the convention of Wigner D matrix and we have explicitly verified it for employed convention given in the Appendix \ref{sec:wignerD}. }.  
        
           Pairs  of $D$'s are combined  
     \bea
      \label{doubleD}
     &D^{s_1}_{m_{s1}m_{s1}'}(R_a^{-1})  D^{s_2}_{m_{s2}m_{s2}'}(R_a^{-1})&=\sum_{\tilde S,\tilde m_S,m_S'} C^{\tilde S,\tilde m_S}_{s_1m_{s1},s_2m_{s2}}C^{\tilde S, m_S'}_{s_1m_{s1}',s_2m_{s2}'}~D^{\tilde  S}_{\tilde m_S m_S'}(R_a^{-1})\nonumber\\
  &   D^{L}_{m_{L}m_{L}'}(R_a^{-1})  D^{\tilde S}_{\tilde m_{S}m_{S}'}(R_a^{-1})&=\sum_{\tilde J,\tilde m_J,m_J'} C^{\tilde J,\tilde m_J}_{Lm_{L},\tilde S \tilde m_{S}}C^{\tilde J, m_J'}_{Lm_{L}',\tilde Sm_{S}'}~D^{\tilde  J}_{\tilde m_J m_J'}(R_a^{-1})
     \eea
     and pairs of Clebsch-Gordan coefficients with the same subscripts are  summed to get
     \be
     \label{doubleC}
    \sum_{m_{s1},m_{s2}} C^{Sm_S}_{s_1m_{s1},s_2m_{s2}}  C^{\tilde S,\tilde m_S}_{s_1m_{s1},s_2m_{s2}}=\delta_{m_S,\tilde m_S}\delta_{S,\tilde S}\ ,\quad 
   \sum_{m_L,m_S } C^{Jm_J}_{Lm_L,Sm_S} C^{\tilde J,\tilde m_J}_{Lm_{L},S m_{S}}=\delta_{m_J,\tilde m_J}\delta_{J,\tilde J}\;. 
        \ee
     Inserting all this to (\ref{10}) one has
     \bea
&&R_aO^{J,m_J,L,S}R_a^{-1}=\nonumber\\
&&=\sum_{m_J'}D_{m_Jm_J'}^J(R_a^{-1}) \hspace*{-12pt} \sum_{m_L',m_S',m_{s1}',m_{s2}'} C^{Jm_J'}_{Lm_L',Sm_S'}C^{Sm_S'}_{s_1m_{s1}',s_2m_{s2}'} \sum_{R'\in O^{(2)}} Y^*_{Lm_L'}(\hat{R'p}) H_{m_{s1}'}(R'p)H_{m_{s2}'}(-R'p)\nonumber\\
&&=\sum_{m_J'}D_{m_Jm_J'}^J(R_a^{-1}) O^{J,m_J',L,S}
\eea
as needed to verify (\ref{4}).  In the last step $R\in O^{(2)}$  was replaced by back to $R\in O$ in (\ref{O_LS}) as both have the same effect on $p$.
  
  \section{Technicalities}\label{sec:technicalities}
  
  \subsection{Wigner D-matrices }\label{sec:wignerD}
  
We employ the conventional definition of  Wigner D-matrix,  which renders (\ref{1}) and is used for example in \cite{Jacob:1959at,pdg14}. That convention differs from the definition in Wigner's book \cite{Wigner:groups}, which is incorporated in Wolfram's  \textsc{Mathematica} \cite{Mathematica10.4}\footnote{In  \textsc{Mathematica} $\texttt{WignerD}[\{j,m,m'\},\alpha,\beta,\gamma]=e^{im\alpha+im^\prime\gamma}\texttt{WignerD}[\{j,m,m'\},\beta]$, while the exponent has different sign for a more conventional definition employed here.}.  The employed $D^j$ for the rotation $R$  that  is consistent with  (\ref{1})  is 
      \begin{align}
  \label{wignerD}
&D^j_{ m, m'}[R^\omega_{ \alpha \beta \gamma}] = F \cdot
  \texttt{WignerD}[\{j, m, m'\}, -\alpha,-\beta,-\gamma],\\
   &F=\left\{
                \begin{array}{ll}
                 1\ :\  j=\mathrm{integer}\ \\ 
                  \pm 1: \ j=\mathrm{halfinteger;\  choice\ of\ sign\ in \ paragraph\ below,\ F(\omega+2\pi )=-F(\omega)~.}   \nonumber \\         
                \end{array}
              \right. 
              \end{align}
 \texttt{WignerD} denotes the \textsc{Mathematica} function and $-\pi \leq\omega<3\pi$ is the angle of rotation $R$ around $\vec n$ in positive direction. The  rotations for $\pi\leq\omega<3\pi$ are present only in the double cover group. 
 
 The  Euler angles of $R$ 
 can be obtained with  \textsc{Mathematica} as $\{\alpha,\beta,\gamma\}=\texttt{EulerAngles}[T]$ with $T=\exp(-i\vec n \vec J \omega)$
  and $(J_k)_{ij}=-i\epsilon_{ijk}$ (\ref{7}). They are the same for $\omega$ and $\omega+2\pi$, and the only difference between these two rotations is given by the factor $F$.  The sign of $F=\pm 1$ for half-integer $j$ and $\pi\leq \omega<\pi$ can be arbitrarily chosen for rotations  $i=1,..,24$  in Table A.1 of  \cite{Bernard:2008ax}: we choose it so that  $D^{1/2}=\exp(-\tfrac{i}{2} \vec n \vec \sigma  \omega)$, which gives $F=-1$ for $i=2,4,12,16,18,20,21,23,24$ and $F=1$ for remaining fifteen. The sign of next 24 elements of $O^2$ with $ \pi\leq \omega<3\pi $ is fixed by  $F(\omega+2\pi)=-F(\omega)$.  The resulting operators do not depend on  the choice of the sign in $F$, except for irrelevant multiplicative factor \footnote{Consider for example   $NV$ helicity operators (\ref{O_helicity}): each term of sum in $R$ contains $F^2=1$, where   one  $F=\pm 1$ is present in  $D^J$ and the same $F$ is present in the rotation of $N$ (both related to half-integer spin).}.
     
  \subsection{Transformations of basis vectors and components}\label{sec:components}
  
 Let the hadron operator $H$ be a linear combination of operators $H_m$ ($m$ can stand for $m_s$ or $i=x,y,z$)   
 \be
  H=\sum c_m H_m=\begin{pmatrix} c_{1} \\ c_{2} \\ ..\end{pmatrix}\;,
  \ee
  where $H_m$ are the basis vectors and $c_m$ the coefficients. If the basis vectors $H_m$ transform 
  according to
  \be
RH_mR^{-1}=\sum_{\tilde m} M_{\tilde m m} H_{\tilde m}
  \ee 
(which can be deduced from (\ref{1},\ref{7})) then   
\be
  RHR^{-1}=\sum_m c_m ~RH_m R^{-1}=\sum_{m,\tilde m} c_m M_{\tilde m m}H_{\tilde m}=\sum_{\tilde m} c^\prime_{\tilde m} H_{\tilde m}
  \ee
  and the coefficients $c$ transform like
  \be
  c^\prime_{\tilde m}=\sum_{m} M_{\tilde m m}c_m \;.
 \ee

  \subsection{ Two bases for vectors }\label{sec:vectors}

  The vector states $V^\dagger|0\rangle=|V\rangle= A_x |V_x\rangle +A_y |V_y\rangle+A_z |V_z\rangle$ with coefficients $A$  \footnote{Coefficients are in agreement with $Y_{11}\propto  -x-iy$, $Y_{10} \propto z$, $Y_{1-1}\propto x-iy$.}
 \begin{equation}
\label{v}
\begin{pmatrix}A_x\\A_y\\A_z\\\end{pmatrix}_{m_J=1}=\tfrac{1}{\sqrt{2}}\begin{pmatrix}-1\\-i\\0\\\end{pmatrix}\ , \qquad \begin{pmatrix}A_x\\A_y\\A_z\\\end{pmatrix}_{m_J=0}=\begin{pmatrix}0\\0\\1\\\end{pmatrix}\ , \qquad \begin{pmatrix}A_x\\A_y\\A_z\\\end{pmatrix}_{m_J=-1}=\tfrac{1}{\sqrt{2}}\begin{pmatrix}1\\-i\\0\\\end{pmatrix}
\end{equation}
have good $S_z$ at rest, where $(S_k)_{ij}=-i\epsilon_{ijk}$
  \begin{equation}
S_z \begin{pmatrix}A_x\\A_y\\A_z\\\end{pmatrix}_{m_J}= -i \begin{pmatrix}0 & 1 & 0 \\ -1 & 0 & 0 \\ 0 & 0 & 0\\\end{pmatrix}  \begin{pmatrix}A_x\\A_y\\A_z\\\end{pmatrix}_{m_J} = m_J \begin{pmatrix}A_x\\A_y\\A_z\\\end{pmatrix}\;.
\end{equation} 
   The annihilation operators $V$ (\ref{6})  are obtained by hermitian conjugation of  $V^\dagger$, so annihilation operators have complex-conjugated coefficients  with respect to the states above.

  \subsection{Subduction matrices ${\cal S}$  }\label{sec:rows}
  
The subduction coefficients in (\ref{14}) depend on the conventions for the rows of the irrep.
  Our conventions  agree with Appendix A of \cite{Bernard:2008ax} for all irreps. This also agrees with rows  \cite{Dudek:2010wm,Edwards:2011jj}  for all irreps except  for $T_{1,2}$.     Our rows are proportional to the entries  $(x,y,z)$ and $(yz,zx,xy)$ for $T_1$ and $T_2$, respectively, while they are related to $(Y_{1,1},Y_{1,0},Y_{1,-1})$ and $(Y_{2,1},\tfrac{1}{2}(Y_{2,2}-Y_{2,-2}),Y_{2,1})$ in \cite{Basak:2005ir,Dudek:2010wm,Edwards:2011jj}. We employ all subduction matrices ${\cal S}$  from \cite{Dudek:2010wm,Edwards:2011jj} except for $T_{1,2}$ which are  deduced from \cite{Edwards:2011jj}:
       \be
(J=1)\to T_1: \\
\begin{array}{c|ccc}
r \backslash m_J & 1 & 0 & -1\\ 
\hline
1& -\frac{1}{\sqrt{2}} & 0 & \frac{1}{\sqrt{2}} \\
2& \frac{1}{\sqrt{2}} & 0 & \frac{1}{\sqrt{2}} \\
3& 0 & 1 & 0 \\
\end{array}~,
\qquad
  (J=2)\to T_2: 
\begin{array}{c|ccccc}
r \backslash m_J & 2 & 1 & 0 & -1 & -2\\ 
\hline
1&	0 & \frac{1}{\sqrt{2}} & 0 & \frac{1}{\sqrt{2}} & 0 \\
2&	0 & \frac{1}{\sqrt{2}} & 0 & -\frac{1}{\sqrt{2}} & 0 \\
3&	\frac{1}{\sqrt{2}} & 0 & 0 & 0 & -\frac{1}{\sqrt{2}} \\
\end{array}~.\nonumber
\ee
 
\section{Expressions for two-hadron operators }\label{sec:HH}

\subsection{PN operators}
\subsubsection{$|p|=0$} 
 \bea
 && G_1^-:\\
 &&O_{G_1^-,r=1}= N_{\frac{1}{2}}(0)\text{P}(0)\qquad\qquad\qquad\qquad\qquad\qquad\qquad\qquad\qquad\qquad\qquad\qquad\qquad  ~
 \nonumber\\
 && O_{G_1^-,r=1}^{[J=\frac{1}{2},L=0,S=\frac{1}{2} ]}= O_{G_1^-,r=1} \nonumber
 \eea
Operators for other irreps are equal to $0$. 
\subsubsection{$|p|=1$}
\bea
&&G_1^ + :\\
&& 
O_{G_1^+,r=1}= 
N_{-\frac{1}{2}}(-e_x) \text{P}(e_x)-N_{-\frac{1}{2}}(e_x) \text{P}(-e_x)-i N_{-\frac{1}{2}}(-e_y) \text{P}(e_y)+i N_{-\frac{1}{2}}(e_y) \text{P}(-e_y) \nnn
&&\qquad \qquad + N_{\frac{1}{2}}(-e_z) \text{P}(e_z)-N_{\frac{1}{2}}(e_z) \text{P}(-e_z) \nnn
&& ~ \nnn
&& O_{G_1^+,r=1}^{[J=\frac{1}{2},m_{J}=\frac{1}{2} ,P=+,\lambda_N=\pm \frac{1}{2},\lambda_P=0]}=O_{G_1^-,r=1}^{[J=\frac{1}{2},m_{J}=\frac{1}{2},L=1,S=\frac{1}{2}]}=  O_{G_1^+,r=1}  \nm
~\nm
O_{\Gdp}=O_{\Gdm}=0   \\
~\nm
 G_1^-: \\ 
&& O_{G_1^-,r=1}=N_{\frac{1}{2}}(-e_x) \text{P}(e_x)+N_{\frac{1}{2}}(e_x) \text{P}(-e_x)+N_{\frac{1}{2}}(-e_y) \text{P}(e_y)+N_{\frac{1}{2}}(e_y)  \text{P}(-e_y) \nonumber \\
&& \qquad \qquad  +  N_{\frac{1}{2}}(-e_z) \text{P}(e_z)+N_{\frac{1}{2}}(e_z) \text{P}(-e_z) \nonumber \\
&& ~ \nonumber \\
&& O_{G_1^-,r=1}^{[J=\frac{1}{2},m_{J}=\frac{1}{2} ,P=-,\lambda_N=\pm \frac{1}{2},\lambda_P=0]}=O_{G_1^-,r=1}^{[J=\frac{1}{2},m_{J}=\frac{1}{2},L=0,S=\frac{1}{2} ]}= O_{G_1^-,r=1} \nn
\eea
\bea
&& \Hp \\
&& 
O_{H^+,r=1}=-i N_{\frac{1}{2}}(-e_x) \text{P}(e_x)+i N_{\frac{1}{2}}(e_x) \text{P}(-e_x)-N_{\frac{1}{2}}(-e_y) \text{P}(e_y)+N_{\frac{1}{2}}(e_y) \text{P}(-e_y) \nnn
&& ~ \nnn
&& O_{H^+,r=1}^{[J=\frac{3}{2},m_{J}=\frac{3}{2} ,P=+, \lambda_N=\pm \frac{1}{2}, \lambda_P=0]}=O_{H^+,r=1}^{[J=\frac{5}{2},P=+, \lambda_N=\pm \frac{1}{2}, \lambda_P=0]}= O_{H^+,r=1} \nm
O_{H^+,r=1}^{[J=\frac{3}{2},m_{J}=\frac{3}{2},L=1,S=\frac{1}{2}]}=O_{H^+,r=1}^{[J=\frac{5}{2},L=3,S=\frac{1}{2}]}= O_{H^+,r=1}  \nm
~\nm
\Hm: \\
&& O_{H^-,r=1}=N_{-\frac{1}{2}}(-e_x) \text{P}(e_x)+N_{-\frac{1}{2}}(e_x) \text{P}(-e_x)- N_{-\frac{1}{2}}(-e_y) \text{P}(e_y)-N_{-\frac{1}{2}}(e_y) \text{P}(-e_y) \nnn
&& ~ \nnn
&& O_{H^-,r=1}^{[J=\frac{3}{2},m_{J}=\frac{3}{2} ,P=-, \lambda_N=\pm \frac{1}{2}, \lambda_P=0]}=O_{H^-,r=1}^{[J=\frac{5}{2} ,P=-, \lambda_N=\pm \frac{1}{2}, \lambda_P=0]}= O_{H^-,r=1}  \nnn
&& O_{H^-,r=1}^{[J=\frac{3}{2},m_{J}=\frac{3}{2},L=2,S=\frac{1}{2}]}= O_{H^-,r=1}^{[J=\frac{5}{2},L=2,S=\frac{1}{2}]}=O_{H^-,r=1} \nn
\eea
The $O_{G_1^+}$ for $|p|=1$ was more recently used to simulate the $N\pi$ scattering in the $J^P=1/2^+$ channel  \cite{Lang:2016hnn}, where the Roper resonance  was found in experiment. 

\subsection{$NV$ operators} 
\subsubsection{$|p|=0$}
 \bea
 && G_1^-:\\
 &&O_{G_1^-,r=1}=N_{\frac{1}{2}}(0) V_z(0)+N_{-\frac{1}{2}}(0) \left(V_x(0)-i V_y(0)\right)\qquad\qquad\qquad\qquad\qquad\qquad  ~
 \nonumber\\
 && O_{G_1^-,r=1}^{[J=\frac{1}{2},m_{J}=\frac{1}{2},L=0,S=\frac{1}{2} ]}= O_{G_1^-,r=1} \nonumber \\
 &&~ \nonumber\\
 && H^-:\\
 &&O_{H^-,r=1}=N_{\frac{1}{2}}(0) \left(V_x(0)-i V_y(0)\right)\nonumber\\
 && O_{H^-,r=1}^{[J=\frac{3}{2},L=0,S=\frac{3}{2} ]}= O_{H^-,r=1} \nonumber
 \eea
Operators for other irreps are equal to $0$. 
\subsubsection{$|p|=1$}
\bea 
&& 
 \Gep: \\
&& 
O_{G_1^+,r=1,n=1}=N_{\frac{1}{2}}(e_x) V_y(-e_x)-N_{\frac{1}{2}}(-e_x) V_y(e_x)+i N_{-\frac{1}{2}}(e_x) V_z(-e_x)-i N_{-\frac{1}{2}}(-e_x) V_z(e_x)\nnn
&& \qquad \qquad \qquad - N_{\frac{1}{2}}(e_y) V_x(-e_y)+N_{\frac{1}{2}}(-e_y) V_x(e_y)+N_{-\frac{1}{2}}(e_y) V_z(-e_y)-N_{-\frac{1}{2}}(-e_y) V_z(e_y)\nnnza \qquad
- i N_{-\frac{1}{2}}(e_z) V_x(-e_z)+i N_{-\frac{1}{2}}(-e_z) V_x(e_z)-N_{-\frac{1}{2}}(e_z) V_y(-e_z)+N_{-\frac{1}{2}}(-e_z) V_y(e_z) \nnn
&& 
O_{G_1^+,r=1,n=2}=N_{\frac{1}{2}}(e_x) V_x(-e_x)-N_{\frac{1}{2}}(-e_x) V_x(e_x)+N_{\frac{1}{2}}(e_y) V_y(-e_y)-N_{\frac{1}{2}}(-e_y) V_y(e_y)\nnnza \qquad
+ N_{\frac{1}{2}}(e_z) V_z(-e_z)-N_{\frac{1}{2}}(-e_z) V_z(e_z) \nnn
&& O_{G_1^+,r=1}^{[J=\frac{1}{2},P=+,\lambda_N= \frac{1}{2},\lambda_V=1]}= O_{G_1^+,r=1}^{[J=\frac{1}{2},P=+,\lambda_N= - \frac{1}{2},\lambda_V=-1]}= O_{G_1^+,r=1,n=1} \nnn
&& O_{G_1^+,r=1}^{[J=\frac{1}{2},P=-,\lambda_N= \pm \frac{1}{2},\lambda_V= 0]}= O_{G_1^+,r=1,n=2} \nnn
&& O_{G_1^+,r=1}^{[J=\frac{1}{2},L=1,S=\frac{1}{2}]}= -i  O_{G_1^+,r=1,n=1} - O_{G_1^+,r=1,n=2} \nnn
&& O_{G_1^+,r=1}^{[J=\frac{1}{2},L=1,S=\frac{3}{2}]}=  -i  O_{G_1^+,r=1,n=1} + 2 ~  O_{G_1^+,r=1,n=2} \nonumber
\eea
\bea
&&\Gem:\\
&& 
O_{G_1^-,r=1,n=1}=N_{-\frac{1}{2}}(e_x) V_y(-e_x)+N_{-\frac{1}{2}}(-e_x) V_y(e_x)+i N_{\frac{1}{2}}(e_x) V_z(-e_x)+i N_{\frac{1}{2}}(-e_x) V_z(e_x)\nnn
&&
 \qquad \qquad \qquad +  i N_{-\frac{1}{2}}(e_y) V_x(-e_y)+ i N_{-\frac{1}{2}}(-e_y) V_x(e_y)+i N_{\frac{1}{2}}(e_y) V_z(-e_y)+i N_{\frac{1}{2}}(-e_y) V_z(e_y)\nnn
&&
\qquad \qquad \qquad +  i N_{-\frac{1}{2}}(e_z) V_x(-e_z)+i N_{-\frac{1}{2}}(-e_z) V_x(e_z)+N_{-\frac{1}{2}}(e_z) V_y(-e_z)+N_{-\frac{1}{2}}(-e_z) V_y(e_z) \nnn 
&&
O_{G_1^-,r=1,n=2}=N_{-\frac{1}{2}}(e_x) V_x(-e_x)+N_{-\frac{1}{2}}(-e_x) V_x(e_x)-i N_{-\frac{1}{2}}(e_y) V_y(-e_y)-i N_{-\frac{1}{2}}(-e_y) V_y(e_y)\nnn
&&\qquad \qquad \qquad + N_{\frac{1}{2}}(e_z) V_z(-e_z)+N_{\frac{1}{2}}(-e_z) V_z(e_z) \nnn
&& ~ \nnn
&&O_{G_1^-,r=1}^{[J=\frac{1}{2},P=- ,\lambda_N= - \frac{1}{2},\lambda_V=-1]}=
O_{G_1^-,r=1}^{[J=\frac{1}{2},P=- ,\lambda_N=  \frac{1}{2},\lambda_V=1]}=  O_{G_1^-,r=1,n=1} \nnn
&& O_{G_1^-,r=1}^{[J=\frac{1}{2},P=-,\lambda_N= \pm \frac{1}{2},\lambda_V= 0]}= 
 O_{G_1^-,r=1,n=2} \nnn
&& O_{G_1^-,r=1}^{[J=\frac{1}{2},L=0,S=\frac{1}{2}]}= - i  O_{G_1^-,r=1,n=1} + O_{G_1^-,r=1,n=2} \nnn
&& O_{G_1^-,r=1}^{[J=\frac{1}{2},L=2,S=\frac{3}{2}]}= -i O_{G_1^-,r=1,n=1}- 2 ~ O_{G_1^-,r=1,n=2} 
\nn
\eea
\bea
&&
\Gdp: \\
&& 
O_{G_2^+,r=1}=N_{\frac{1}{2}}(e_x) V_y(-e_x)-N_{\frac{1}{2}}(-e_x) V_y(e_x)-i N_{-\frac{1}{2}}(e_x) V_z(-e_x)+i N_{-\frac{1}{2}}(-e_x) V_z(e_x) \nnn
&& 
\qquad \qquad + N_{\frac{1}{2}}(e_y) V_x(-e_y)-N_{\frac{1}{2}}(-e_y) V_x(e_y)+N_{-\frac{1}{2}}(e_y) V_z(-e_y)-N_{-\frac{1}{2}}(-e_y) V_z(e_y)\nnn
&& 
 \qquad \qquad-i N_{-\frac{1}{2}}(e_z) V_x(-e_z)+i N_{-\frac{1}{2}}(-e_z) V_x(e_z)+N_{-\frac{1}{2}}(e_z) V_y(-e_z)-N_{-\frac{1}{2}}(-e_z) V_y(e_z) \nnn
&& O_{G_2^+,r=1}^{[J=\frac{5}{2},P=+,\lambda_N= - \frac{1}{2},\lambda_V=1]}=  
O_{G_2^+,r=1}^{[J=\frac{5}{2},P=+,\lambda_N=  \frac{1}{2},\lambda_V=-1]}=O_{G_2^+,r=1}^{[J=\frac{5}{2},L=1,S=\frac{3}{2}]}=  O_{G_2^+,r=1} \nonumber\\ 
~&&\nonumber
\eea
\bea
&&\Gdm: \\
&&
O_{G_2^-,r=1}=N_{-\frac{1}{2}}(e_x) V_y(-e_x)+N_{-\frac{1}{2}}(-e_x) V_y(e_x)-i N_{\frac{1}{2}}(e_x) V_z(-e_x)-i N_{\frac{1}{2}}(-e_x) V_z(e_x)\nnn
&& \qquad \qquad -  i N_{-\frac{1}{2}}(e_y) V_x(-e_y)-i N_{-\frac{1}{2}}(-e_y) V_x(e_y)+i N_{\frac{1}{2}}(e_y) V_z(-e_y)+i N_{\frac{1}{2}}(-e_y) V_z(e_y)\nnn
&& \qquad\qquad+i N_{-\frac{1}{2}}(e_z) V_x(-e_z)+i N_{-\frac{1}{2}}(-e_z) V_x(e_z)-N_{-\frac{1}{2}}(e_z) V_y(-e_z)-N_{-\frac{1}{2}}(-e_z) V_y(e_z) \nnn
&& ~ \nnn
&& O_{G_2^-,r=1}^{[J=\frac{5}{2},P=-,\lambda_N= - \frac{1}{2},\lambda_V=1]}= 
O_{G_2^-,r=1}^{[J=\frac{5}{2},P=-,\lambda_N= \frac{1}{2},\lambda_V=-1]}=  O_{G_2^-,r=1} \nnn
&& O_{G_2^-,r=1}^{[J=\frac{5}{2},L=2,S=\frac{3}{2}]}= O_{G_2^-,r=1}^{[J=\frac{5}{2},L=4,S=\frac{3}{2}]}= O_{G_2^-,r=1} 
\nn
\eea
\bea
&& \Hp : \\
&& O_{H^+,r=1,n=1}=N_{-\frac{1}{2}}(e_x) V_x(-e_x)-N_{-\frac{1}{2}}(-e_x) V_x(e_x)-N_{-\frac{1}{2}}(e_y) V_y(-e_y)+N_{-\frac{1}{2}}(-e_y) V_y(e_y) \nnn
&& O_{H^+,r=1,n=2}=-N_{-\frac{1}{2}}(e_x) V_y(-e_x)+N_{-\frac{1}{2}}(-e_x) V_y(e_x)+i N_{\frac{1}{2}}(e_x) V_z(-e_x)-i N_{\frac{1}{2}}(-e_x) V_z(e_x)\nnn
&& \qquad \qquad \qquad -N_{-\frac{1}{2}}(e_y) V_x(-e_y)+N_{-\frac{1}{2}}(-e_y) V_x(e_y)+N_{\frac{1}{2}}(e_y) V_z(-e_y)-N_{\frac{1}{2}}(-e_y) V_z(e_y)\nnn
&& \qquad \qquad \qquad -2 i N_{\frac{1}{2}}(e_z) \left(V_x(-e_z)-i V_y(-e_z)\right)+2 N_{\frac{1}{2}}(-e_z) \left(V_y(e_z)+i V_x(e_z)\right) \nnn
&&O_{H^+,r=1,n=3}=
-N_{-\frac{1}{2}}(e_x) V_y(-e_x)+N_{-\frac{1}{2}}(-e_x) V_y(e_x)-2 i N_{\frac{1}{2}}(e_x) V_z(-e_x)+2 i N_{\frac{1}{2}}(-e_x) V_z(e_x)\nnn
&& \qquad \qquad \qquad -N_{-\frac{1}{2}}(e_y) V_x(-e_y)+N_{-\frac{1}{2}}(-e_y) V_x(e_y)-2 N_{\frac{1}{2}}(e_y) V_z(-e_y)+2 N_{\frac{1}{2}}(-e_y) V_z(e_y)\nnn
&& \qquad \qquad \qquad +N_{\frac{1}{2}}(e_z)\left(V_y(-e_z)+i V_x(-e_z)\right)-i N_{\frac{1}{2}}(-e_z) \left(V_x(e_z)-i V_y(e_z)\right) \nnn
&& O_{H^+,r=1}^{[J=\frac{3}{2},P=+,\lambda_N= \pm  \frac{1}{2},\lambda_V= 0]}= O_{H^+,r=1}^{[J=\frac{5}{2},P=+,\lambda_N= \pm  \frac{1}{2},\lambda_V= 0]}= O_{H^+,r=1,n=1} \nnn
&& O_{H^+,r=1}^{[J=\frac{3}{2},P=+,\lambda_N= -\frac{1}{2},\lambda_V=  1]}=
O_{H^+,r=1}^{[J=\frac{3}{2},P=+,\lambda_N=  \frac{1}{2},\lambda_V=  -1]}=\nm
\qquad ~ ~  ~ ~
=O_{H^+,r=1}^{[J=\frac{5}{2},P=+,\lambda_N= -\frac{1}{2},\lambda_V= 1]}=
O_{H^+,r=1}^{[J=\frac{5}{2},P=+,\lambda_N=  \frac{1}{2},\lambda_V= - 1]}=
O_{H^+,r=1,n=2} \nnn
&&O_{H^+,r=1}^{[J=\frac{3}{2},P=+,\lambda_N= -\frac{1}{2},\lambda_V= - 1]}= 
O_{H^+,r=1}^{[J=\frac{3}{2},P=+,\lambda_N=\frac{1}{2},\lambda_V= 1]}=\nm
\qquad  ~ ~  ~ ~
=O_{H^+,r=1}^{[J=\frac{5}{2},P=+,\lambda_N=  \frac{1}{2},\lambda_V=1]}= 
O_{H^+,r=1}^{[J=\frac{5}{2},P=+,\lambda_N=-\frac{1}{2},\lambda_V=-1]}=
O_{H^+,r=1,n=2} + 2~ O_{H^+,r=1,n=3} \nnn 
&& O_{H^+,r=1}^{[J=\frac{3}{2},L=1,S=\frac{3}{2}]}=3~   O_{H^+,r=1,n=1} +i  (4~ O_{H^+,r=1,n=2} - O_{H^+,r=1,n=3}) \nnn
&&
O_{H^+,r=1}^{[J=\frac{3}{2},L=1,S=\frac{1}{2}]}=O_{H^+,r=1}^{[J=\frac{5}{2},L=3,S=\frac{1}{2}]}=3~  O_{H^+,r=1,n=1} +i (O_{H^+,r=1,n=2}+ 2 ~ O_{H^+,r=1,n=3})\nnn
&& 
O_{H^+,r=1}^{[J=\frac{3}{2},L=3,S=   \frac{3}{2}]}=O_{H^+,r=1}^{[J=\frac{5}{2},L=1,S=\frac{3}{2}]}= 3 ~ O_{H^+,r=1,n=1} -i  ( O_{H^+,r=1,n=2} + O_{H^+,r=1,n=3}) \nm
O_{H^+,r=1}^{[J=\frac{5}{2},L=3,S=\frac{3}{2}]}=12~O_{H^+,r=1,n=1} +i  (O_{H^+,r=1,n=2} -4~O_{H^+,r=1,n=3}) \nn
\eea 
\bea
&& \Hm : \\
&& 
O_{H^-,r=1,n=1}=i N_{\frac{1}{2}}(e_x) V_x(-e_x)+i N_{\frac{1}{2}}(-e_x) V_x(e_x)+N_{\frac{1}{2}}(e_y) V_y(-e_y)+N_{\frac{1}{2}}(-e_y) V_y(e_y) \nnn
&& 
O_{H^-,r=1,n=2}=-i N_{\frac{1}{2}}(e_x) V_y(-e_x)-i N_{\frac{1}{2}}(-e_x) V_y(e_x)-N_{-\frac{1}{2}}(e_x) V_z(-e_x)-N_{-\frac{1}{2}}(-e_x) V_z(e_x)\nnn
&& \qquad \qquad \qquad+  N_{\frac{1}{2}}(e_y) V_x(-e_y)+N_{\frac{1}{2}}(-e_y) V_x(e_y)+N_{-\frac{1}{2}}(e_y) V_z(-e_y)+N_{-\frac{1}{2}}(-e_y) V_z(e_y)\nnn
&& \qquad \qquad \qquad+ 2 N_{\frac{1}{2}}(e_z) \left(V_x(-e_z)-i V_y(-e_z)\right)+2 N_{\frac{1}{2}}(-e_z) \left(V_x(e_z)-i V_y(e_z)\right) \nnn
&&
O_{H^-,r=1,n=3}=i N_{\frac{1}{2}}(e_x) V_y(-e_x)+i N_{\frac{1}{2}}(-e_x) V_y(e_x)-2 N_{-\frac{1}{2}}(e_x) V_z(-e_x)-2 N_{-\frac{1}{2}}(-e_x) V_z(e_x)\nnn
&& \qquad \qquad \qquad- N_{\frac{1}{2}}(e_y) V_x(-e_y)-N_{\frac{1}{2}}(-e_y) V_x(e_y)+2 N_{-\frac{1}{2}}(e_y) V_z(-e_y)+2 N_{-\frac{1}{2}}(-e_y) V_z(e_y)\nnn
&& \qquad \qquad \qquad +  N_{\frac{1}{2}}(e_z) \left(V_x(-e_z)-i V_y(-e_z)\right)+N_{\frac{1}{2}}(-e_z) \left(V_x(e_z)-i V_y(e_z)\right) \nm
O_{H^-,r=1}^{[J=\frac{3}{2},P=-,\lambda_N= \pm \frac{1}{2},\lambda_V= 0]}=O_{H^-,r=1}^{[J=\frac{5}{2},P=-,\lambda_N= \pm \frac{1}{2},\lambda_V= 0]}=   O_{H^-,r=1,n=1} \nnn
&& 
O_{H^-,r=1}^{[J=\frac{3}{2},P=-,\lambda_N=  \frac{1}{2},\lambda_V= - 1]}= 
O_{H^-,r=1}^{[J=\frac{3}{2},P=-,\lambda_N= -\frac{1}{2},\lambda_V=  1]}= \nm
\qquad ~ ~  ~ ~
=O_{H^-,r=1}^{[J=\frac{5}{2},P=-,\lambda_N=  \frac{1}{2},\lambda_V= - 1]}= O_{H^-,r=1}^{[J=\frac{5}{2},P=-,\lambda_N= -\frac{1}{2},\lambda_V=  1]}=
O_{H^-,r=1,n=2} \nnn
&& O_{H^-,r=1}^{[J=\frac{3}{2},P=-,\lambda_N= \frac{1}{2},\lambda_V= 1]}=
O_{H^-,r=1}^{[J=\frac{3}{2},P=-,\lambda_N= - \frac{1}{2},\lambda_V= -1]}= \nm
\qquad~ ~  ~ ~  =O_{H^-,r=1}^{[J=\frac{5}{2},P=-,\lambda_N= \frac{1}{2},\lambda_V= 1]}=
O_{H^-,r=1}^{[J=\frac{5}{2},P=-,\lambda_N= - \frac{1}{2},\lambda_V= -1]}=
O_{H^-,r=1,n=2}- 2~  O_{H^-,r=1,n=3} \nnn
&&
O_{H^-,r=1}^{[J=\frac{3}{2},L=0,S=\frac{3}{2}]}=  -3i~ O_{H^-,r=1,n=1}+2~ O_{H^-,r=1,n=2} - O_{H^-,r=1,n=3} \nnn
&& O_{H^-,r=1}^{[J=\frac{3}{2},L=2,S=  \frac{3}{2}]}=O_{H^-,r=1}^{[J=\frac{5}{2},L=2,S=\frac{3}{2}]}= 3i~  O_{H^-,r=1,n=1} + O_{H^-,r=1,n=2} + O_{H^-,r=1,n=3} \nnn
&&
O_{H^-,r=1}^{[J=\frac{3}{2},L=2,S=   \frac{1}{2}]}=  O_{H^-,r=1}^{[J=\frac{5}{2},L=2,S=\frac{1}{2}]}= 3 i~ O_{H^-,r=1,n=1}+  O_{H^-,r=1,n=2} - 2 ~ O_{H^-,r=1,n=3} \nnn
&&
O_{H^-,r=1}^{[J=\frac{5}{2},L=4,S=\frac{3}{2}]}= 12i ~ O_{H^-,r=1,n=1}- 3 ~O_{H^-,r=1,n=2} + 4 ~ O_{H^-,r=1,n=3} \nn
~ \nn
\eea

\subsection{$NN$ operators}
\subsubsection{$|p|=0$}
 \bea
 && A_1^+:\\
 &&O_{A_1^+,r=1}= 
\text{N}_{\frac{1}{2}}(0) \text{N'}_{-\frac{1}{2}}(0)-\text{N}_{-\frac{1}{2}}(0) \text{N'}_{\frac{1}{2}}(0)\qquad\qquad\qquad\qquad\qquad\qquad\qquad\qquad\qquad  ~
 \nonumber\\
 && O_{A_1^+,r=1}^{[J=0,m_{J}=0,L=0,S=0]}= O_{A_1^+,r=1} \nonumber \\
 &&~ \nonumber\\
 && T_1^+:\\
 &&O_{T_1^+,r=1}=\text{N}_{-\frac{1}{2}}(0) \text{N'}_{-\frac{1}{2}}(0)-\text{N}_{\frac{1}{2}}(0) \text{N'}_{\frac{1}{2}}(0)\nonumber\\
 && O_{T_1^+,r=1}^{[J=1, L=0,S=1 ]}= O_{T_1^+,r=1} \nonumber
 \eea
Operators for other irreps vanish. For two identical fermions $N=N'$, the $O_{T_1^+}={N}_{-\frac{1}{2}}(0) \text{N}_{-\frac{1}{2}}(0)-\text{N}_{\frac{1}{2}}(0) \text{N}_{\frac{1}{2}}(0) =0$ due to anticommutation of nucleon fields. This is also required by the Fermi-Dirac statistics for this case as spin, isospin and spatial parts are all symmetric under exchange of two particles. 
 
\subsubsection{$|p|=1$}
\bea 
&& O_{\Adp}=O_{\Adm}=0 \\
&& ~ \nm
\Aep : \\
&& O_{A_1^+,r=1}=
-\text{N}_{-\frac{1}{2}}(e_x) \text{N'}_{\frac{1}{2}}(-e_x)-\text{N}_{-\frac{1}{2}}(-e_x) \text{N'}_{\frac{1}{2}}(e_x)
+\text{N}_{\frac{1}{2}}(e_x) \text{N'}_{-\frac{1}{2}}(-e_x)+\text{N}_{\frac{1}{2}}(-e_x) \text{N'}_{-\frac{1}{2}}(e_x) \nnn
&& \qquad \qquad-\text{N}_{-\frac{1}{2}}(e_y) \text{N'}_{\frac{1}{2}}(-e_y)-\text{N}_{-\frac{1}{2}}(-e_y) \text{N'}_{\frac{1}{2}}(e_y)
+\text{N}_{\frac{1}{2}}(e_y) \text{N'}_{-\frac{1}{2}}(-e_y)+\text{N}_{\frac{1}{2}}(-e_y) \text{N'}_{-\frac{1}{2}}(e_y) \nnn
&& \qquad \qquad-\text{N}_{-\frac{1}{2}}(e_z) \text{N'}_{\frac{1}{2}}(-e_z)-\text{N}_{-\frac{1}{2}}(-e_z) \text{N'}_{\frac{1}{2}}(e_z)
+\text{N}_{\frac{1}{2}}(e_z) \text{N'}_{-\frac{1}{2}}(-e_z)+\text{N}_{\frac{1}{2}}(-e_z) \text{N'}_{-\frac{1}{2}}(e_z) \nnn
&&
O_{A_1^+,r=1}^{[J=0 ,P=+, \lambda_{N}=\frac{1}{2} ,\lambda_{N'}=\frac{1}{2}]}=O_{A_1^+,r=1}^{[J=0 ,P=+, \lambda_{N}=-\frac{1}{2} ,\lambda_{N'}=-\frac{1}{2}]}=O_{A_1^+,r=1}^{[J=0,L=0,S=0]}= O_{A_1^+,r=1} \nonumber  \\ 
&& ~ \nnn
 && A_1^-:\\
 && O_{A_1^-,r=1}:=
 -\text{N}_{\frac{1}{2}}(e_x)\text{N'}_{\frac{1}{2}}(-e_x)+\text{N}_{\frac{1}{2}}(-e_x)\text{N'}_{\frac{1}{2}}(e_x)  +\text{N}_{-\frac{1}{2}}(e_x)\text{N'}_{-\frac{1}{2}}(-e_x)-\text{N}_{-\frac{1}{2}}(-e_x)\text{N'}_{-\frac{1}{2}}(e_x) \nnn
 &&\qquad \qquad \qquad-i \text{N}_{\frac{1}{2}}(e_y) \text{N'}_{\frac{1}{2}}(-e_y)+i \text{N}_{\frac{1}{2}}(-e_y) \text{N'}_{\frac{1}{2}}(e_y)-i \text{N}_{-\frac{1}{2}}(e_y) \text{N'}_{-\frac{1}{2}}(-e_y)+i \text{N}_{-\frac{1}{2}}(-e_y) \text{N'}_{-\frac{1}{2}}(e_y) \nnn
 &&\qquad \qquad \qquad +\text{N}_{-\frac{1}{2}}(e_z) \text{N'}_{\frac{1}{2}}(-e_z)-\text{N}_{-\frac{1}{2}}(-e_z) \text{N'}_{\frac{1}{2}}(e_z)+\text{N}_{\frac{1}{2}}(e_z) \text{N'}_{-\frac{1}{2}}(-e_z)-\text{N}_{\frac{1}{2}}(-e_z) \text{N'}_{-\frac{1}{2}}(e_z) \nnn
&&~ \nonumber \\
&& O_{A_1^-,r=1}^{[J=0 ,P=-, \lambda_{N}=\frac{1}{2} ,\lambda_{N'}=\frac{1}{2}]}=O_{A_1^-,r=1}^{[J=0 ,P=-, \lambda_{N}=-\frac{1}{2} ,\lambda_{N'}=-\frac{1}{2}]}= O_{A_1^-,r=1}^{[J=0,L=1,S=1]}=O_{A_1^-,r=1} \nonumber \\
&&~\nonumber\\
&& \Tep: \\
&&
O_{T_1^+,r=1,n=1}=
-\text{N}_{\frac{1}{2}}(e_x)\text{N'}_{\frac{1}{2}}(-e_x)-\text{N}_{\frac{1}{2}}(-e_x)\text{N'}_{\frac{1}{2}}(e_x)+\text{N}_{-\frac{1}{2}}(e_x)\text{N'}_{-\frac{1}{2}}(-e_x)+\text{N}_{-\frac{1}{2}}(-e_x) \text{N'}_{-\frac{1}{2}}(e_x) \nm
O_{T_1^+,r=1,n=2}=
-\text{N}_{\frac{1}{2}}(e_y) \text{N'}_{\frac{1}{2}}(-e_y)-\text{N}_{\frac{1}{2}}(-e_y) \text{N'}_{\frac{1}{2}}(e_y)+\text{N}_{-\frac{1}{2}}(e_y)\text{N'}_{-\frac{1}{2}}(-e_y)+\text{N}_{-\frac{1}{2}}(-e_y)\text{N'}_{-\frac{1}{2}}(e_y)
\nnnzad
-\text{N}_{\frac{1}{2}}(e_z) \text{N'}_{\frac{1}{2}}(-e_z)-\text{N}_{\frac{1}{2}}(-e_z) \text{N'}_{\frac{1}{2}}(e_z)+\text{N}_{-\frac{1}{2}}(e_z) \text{N'}_{-\frac{1}{2}}(-e_z)+\text{N}_{-\frac{1}{2}}(-e_z) \text{N'}_{-\frac{1}{2}}(e_z) \nm
O_{T_1^+,r=1}^{[J=1,P=+, \lambda_{N}=\frac{1}{2} ,\lambda_{N'}=\frac{1}{2}]}=
 O_{T_1^+,r=1}^{[J=1,P=+, \lambda_{N}=-\frac{1}{2} ,\lambda_{N'}=-\frac{1}{2}]}= 
  O_{T_1^+,r=1, n=1} \nm
O_{T_1^+,r=1}^{[J=1,P=+, \lambda_{N}=-\frac{1}{2} ,\lambda_{N'}=\frac{1}{2}]}=
O_{T_1^+,r=1}^{[J=1,P=+, \lambda_{N}=\frac{1}{2} ,\lambda_{N'}=-\frac{1}{2}]}= 
 O_{T_1^+,r=1, n=2} \nm
O_{T_1^+,r=1}^{[J=1,L=2,S=1]}=2 ~     O_{T_1^+,r=1, n=1} - O_{T_1^+,r=1, n=2}\nm
O_{T_1^+,r=1}^{[J=1,L=0, S=1]}= O_{T_1^+,r=1, n=1} +   O_{T_1^+,r=1, n=2} \nonumber 
\eea
\bea
 &&\Tem: \\
&& O_{T_1^-,r=1,n=1}=
\text{N}_{-\frac{1}{2}}(e_y) \text{N'}_{\frac{1}{2}}(-e_y)-\text{N}_{-\frac{1}{2}}(-e_y)\text{N'}_{\frac{1}{2}}(e_y)
+\text{N}_{\frac{1}{2}}(e_y) \text{N'}_{-\frac{1}{2}}(-e_y)-\text{N}_{\frac{1}{2}}(-e_y)\text{N'}_{-\frac{1}{2}}(e_y)\nnnza  \qquad
+i \text{N}_{\frac{1}{2}}(e_z) \text{N'}_{\frac{1}{2}}(-e_z)-i\text{N}_{\frac{1}{2}}(-e_z)\text{N'}_{\frac{1}{2}}(e_z)
+i \text{N}_{-\frac{1}{2}}(e_z) \text{N'}_{-\frac{1}{2}}(-e_z)-i\text{N}_{-\frac{1}{2}}(-e_z)\text{N'}_{-\frac{1}{2}}(e_z) \nm
O_{T_1^-,r=1,n=2}=-\text{N}_{-\frac{1}{2}}(e_x) \text{N'}_{\frac{1}{2}}(-e_x)+\text{N}_{-\frac{1}{2}}(-e_x) \text{N'}_{\frac{1}{2}}(e_x)+\text{N}_{\frac{1}{2}}(e_x) \text{N'}_{-\frac{1}{2}}(-e_x)-\text{N}_{\frac{1}{2}}(-e_x) \text{N'}_{-\frac{1}{2}}(e_x) \nm
O_{T_1^-,r=1}^{[J=1,P=-, \lambda_{N}=\frac{1}{2} ,\lambda_{N'}=\frac{1}{2}]}=O_{T_1^-,r=1}^{[J=1,P=-, \lambda_{N}=-\frac{1}{2} ,\lambda_{N'}=-\frac{1}{2}]}= O_{T_1^-,r=1,n=2} \nm
O_{T_1^-,r=1}^{[J=1,P=-, \lambda_{N}=\frac{1}{2} ,\lambda_{N'}=-\frac{1}{2}]}=O_{T_1^-,r=1}^{[J=1,P=-, \lambda_{N}=-\frac{1}{2} ,\lambda_{N'}=\frac{1}{2}]}= O_{T_1^-,r=1,n=1} \nm
O_{T_1^-,r=1}^{[J=1,L=1,S=1 ]}= O_{T_1^-,r=1,n=1} \nm
O_{T_1^-,r=1}^{[J=1,L=1, S=0] }=O_{T_1^-,r=1,n=2} \nonumber
\eea
\bea
&& \Tdp \\
&& O_{T_2^+,r=1}=
-\text{N}_{\frac{1}{2}}(e_y)\text{N'}_{\frac{1}{2}}(-e_y)-\text{N}_{\frac{1}{2}}(-e_y)\text{N'}_{\frac{1}{2}}(e_y)
+\text{N}_{-\frac{1}{2}}(e_y)\text{N'}_{-\frac{1}{2}}(-e_y)+\text{N}_{-\frac{1}{2}}(-e_y)\text{N'}_{-\frac{1}{2}}(e_y) \nnnza
+\text{N}_{\frac{1}{2}}(e_z)\text{N'}_{\frac{1}{2}}(-e_z)+\text{N}_{\frac{1}{2}}(-e_z)\text{N'}_{\frac{1}{2}}(e_z)
-\text{N}_{-\frac{1}{2}}(e_z)\text{N'}_{-\frac{1}{2}}(-e_z)-\text{N}_{-\frac{1}{2}}(-e_z)\text{N'}_{-\frac{1}{2}}(e_z) \nm
~ \nm
O_{T_2^+,r=1}^{[J=2,P=+ , \lambda_{N}=\frac{1}{2} ,\lambda_{N'}=-\frac{1}{2}]}=O_{T_2^+,r=1}^{[J=2,P=+ , \lambda_{N}=-\frac{1}{2} ,\lambda_{N'}=\frac{1}{2}]}=O_{T_2^+,r=1}^{[J=2,L=2,S=1]}= O_{T_2^+,r=1} \nm
 ~ \nm
 \Tdm: \\
&&O_{T_2^-,r=1}=i \text{N}_{-\frac{1}{2}}(e_y) \text{N'}_{\frac{1}{2}}(-e_y)-i \text{N}_{-\frac{1}{2}}(-e_y) \text{N'}_{\frac{1}{2}}(e_y)+i \text{N}_{\frac{1}{2}}(e_y) \text{N'}_{-\frac{1}{2}}(-e_y)-i \text{N}_{\frac{1}{2}}(-e_y) \text{N'}_{-\frac{1}{2}}(e_y) \nnn
&& \qquad \qquad +\text{N}_{\frac{1}{2}}(e_z)  \text{N'}_{\frac{1}{2}}(-e_z)-\text{N}_{\frac{1}{2}}(-e_z) \text{N'}_{\frac{1}{2}}(e_z)+\text{N}_{-\frac{1}{2}}(e_z) \text{N'}_{-\frac{1}{2}}(-e_z)-\text{N}_{-\frac{1}{2}}(-e_z) \text{N'}_{-\frac{1}{2}}(e_z) \nnn
&& ~ \nnn
&&O_{T_2^-,r=1}^{[J=2,P=-, \lambda_{N}=\frac{1}{2} ,\lambda_{N'}=-\frac{1}{2}]}=O_{T_2^-,r=1}^{[J=2,P=-, \lambda_{N}=-\frac{1}{2} ,\lambda_{N'}=\frac{1}{2}]}=O_{T_2^-,r=1}^{[J=2,L=1,S=1]}=O_{T_2^-,r=1}^{[J=2,L=3,S=1]}=O_{T_2^-,r=1} \nn
\eea
\bea
&&  \Ep: \\ 
&& O_{E^+,r=1}=
-\text{N}_{-\frac{1}{2}}(e_x) \text{N'}_{\frac{1}{2}}(-e_x)-\text{N}_{-\frac{1}{2}}(-e_x) \text{N'}_{\frac{1}{2}}(e_x)+\text{N}_{\frac{1}{2}}(e_x) \text{N'}_{-\frac{1}{2}}(-e_x)  +\text{N}_{\frac{1}{2}}(-e_x) \text{N'}_{-\frac{1}{2}}(e_x) \nnnza 
-\text{N}_{-\frac{1}{2}}(e_y) \text{N'}_{\frac{1}{2}}(-e_y)-\text{N}_{-\frac{1}{2}}(-e_y) \text{N'}_{\frac{1}{2}}(e_y)+\text{N}_{\frac{1}{2}}(e_y) \text{N'}_{-\frac{1}{2}}(-e_y) +\text{N}_{\frac{1}{2}}(-e_y) \text{N'}_{-\frac{1}{2}}(e_y) \nnnza 
+2 \text{N}_{-\frac{1}{2}}(e_z) \text{N'}_{\frac{1}{2}}(-e_z)+2 \text{N}_{-\frac{1}{2}}(-e_z) \text{N'}_{\frac{1}{2}}(e_z)-2 \text{N}_{\frac{1}{2}}(e_z) \text{N'}_{-\frac{1}{2}}(-e_z)-2 \text{N}_{\frac{1}{2}}(-e_z) \text{N'}_{-\frac{1}{2}}(e_z)\nnn
&& O_{E^+,r=1}^{[J=2,P=+, \lambda_{N}=\frac{1}{2} ,\lambda_{N'}=\frac{1}{2}]}=O_{E^+,r=1}^{[J=2,P=+, \lambda_{N}=-\frac{1}{2} ,\lambda_{N'}=-\frac{1}{2}]}=O_{E^+,r=1}^{[J=2,L=2 ,S=0]}=O_{E^+,r=1} \nm
~ \nm
 \Em : \\ 
&& O_{E^-,r=1}=
-\text{N}_{\frac{1}{2}}(e_x) \text{N'}_{\frac{1}{2}}(-e_x)+\text{N}_{\frac{1}{2}}(-e_x) \text{N'}_{\frac{1}{2}}(e_x)+\text{N}_{-\frac{1}{2}}(e_x) \text{N'}_{-\frac{1}{2}}(-e_x)-\text{N}_{-\frac{1}{2}}(-e_x) \text{N'}_{-\frac{1}{2}}(e_x) \nnn
&& \qquad \qquad  -i \text{N}_{\frac{1}{2}}(e_y) \text{N'}_{\frac{1}{2}}(-e_y)+i \text{N}_{\frac{1}{2}}(-e_y) \text{N'}_{\frac{1}{2}}(e_y)-i \text{N}_{-\frac{1}{2}}(e_y) \text{N'}_{-\frac{1}{2}}(-e_y)+i \text{N}_{-\frac{1}{2}}(-e_y) \text{N'}_{-\frac{1}{2}}(e_y) \nnn
&&\qquad \qquad  -2 \text{N}_{-\frac{1}{2}}(e_z) \text{N'}_{\frac{1}{2}}(-e_z)+2 \text{N}_{-\frac{1}{2}}(-e_z) \text{N'}_{\frac{1}{2}}(e_z)-2 \text{N}_{\frac{1}{2}}(e_z) \text{N'}_{-\frac{1}{2}}(-e_z)+2 \text{N}_{\frac{1}{2}}(-e_z) \text{N'}_{-\frac{1}{2}}(e_z) \nnn
&&O_{E^-,r=1}^{[J=2,P=-,\lambda_{N}=\frac{1}{2} ,\lambda_{N'}=\frac{1}{2}]}=O_{E^-,r=1}^{[J=2,P=-,\lambda_{N}=-\frac{1}{2} ,\lambda_{N'}=-\frac{1}{2}]}=O_{E^-,r=1}^{[J=2,L=1,S=1]}=O_{E^-,r=1}^{[J=2,L=3,S=1]}=O_{E^-,r=1} \nn
\eea
The $L=0$ and $L=2$ partial waves in the deuteron channel ($J^P=1^+$ and $S=1$) could be explored with  $O_{T_1^+}^{[J=1,L=0, S=1]} $ and $O_{T_1^+}^{[J=1,L=2, S=1]} $, respectively.  These handles are expected to be effective since  the physical mixing between $(L=0,S=1)$ and  $(L=2,S=1)$ is small due to the smallness of the tensor force\footnote{The prospects for the lattice extraction of this mixing   has been analytically explored in \cite{Briceno:2013bda}.}.

\section{Relation between partial wave and helicity operators}\label{sec:relation}

The proof of the  relation (\ref{O_rel}) between partial-wave operators (\ref{O_LS}) and helicity operators (\ref{O_helicity}) is presented here.   The Clebsch-Gordan coefficient in the partial wave operator (\ref{O_LS}) can be rewritten  as
\bea
&&C^{Sm_S}_{s_1m_{s1},s_2m_{s2}}= C^{Sm_S}_{s_1m_{s1},s_2m_{s2}} \times \delta(m_S,m_{s1}+m_{s2})=C_{s_1m_{s1},s_2m_{s2}}^{S,m_{s1}+m_{s2}}~, \nonumber
\eea 
 which leads to
\bea
\label{28}
&& O^{|p|,J,m_J,S,L}  =\sum_{R\in O}   (\sum_{m_L,m_S} C^{Jm_J}_{Lm_L,Sm_S}Y^*_{Lm_L}(\widehat{Rp}))  ~ \sum_{m_{s1},m_{s2}}C_{s_1m_{s1},s_2m_{s2}}^{S,m_{s1}+m_{s2}}  H^{(1)}_{m_{s1}}(Rp)H^{(2)}_{m_{s2}}(-Rp)~. ~ \nonumber \\
&& 
\eea
Let us first consider the  term with the sum over $m_{s1}$ and $m_{s2}$. 
The single-hadron operators $H(\pm Rp)$ are expressed in terms of  $H(\pm p_z)$ via $H_{m_s}(R^\prime p_z)=\sum_{\lambda} D_{m_s\lambda}(R^\prime) R^\prime H_{\lambda}(p_z)(R^{\prime})^{-1}$  (\ref{2})\footnote{The summation index is denoted $-\lambda_2$ for the second hadron.} with $Rp=R R_0^p p_z=R^\prime p_z$. Then products of two Wigner D-matrices are reduced via (\ref{doubleD}) and subsequently the product of two CG's via (\ref{doubleC}) leading to  
\bea
\label{17}
&& \sum_{m_{s1},m_{s2}} C_{s_1m_{s1},s_2m_{s2}}^{S,m_{s1}+m_{s2}} H^{(1)}_{m_{s1}}(Rp)H^{(2)}_{m_{s2}}(-Rp)=  \\ 
&& = \sum_{m_{s1},m_{s2}} C_{s_1m_{s1},s_2m_{s2}}^{S,m_{s1}+m_{s2}} \sum_{\lambda_{1},\lambda_{2}} D_{m_{s1},\lambda_1}^{s_1}(R R_0^p)D_{m_{s2},-\lambda_2}^{s_2}(R R_0^p)~(R R_0^p)~ H^{(1)}_{\lambda_{1}}(p_z)H^{(2)}_{-\lambda_{2}}(-p_z)~(R R_0^p)^{-1}  \nonumber \\
&&= \sum_{m_{s1},m_{s2},\lambda_{1},\lambda_{2}}\sum_{\lambda,m_S^{'}}\sum_{\tilde{S}} C_{s_1m_{s1},s_2m_{s2}}^{S,m_{s1}+m_{s2}} C_{s_1m_{s1},s_2m_{s2}}^{\tilde{S},m_{S}^{'}} C_{s_1 \lambda_{1},s_2 -\lambda_{2}}^{\tilde{S},\lambda}  D^{\tilde{S}}_{m_{S}^{'},\lambda}(R R_0^p)~ \nonumber\\
&&  \qquad \qquad  \qquad \qquad \qquad\times\  (R R_0^p)~H^{(1)}_{\lambda_{1}}(p_z)H^{(2)}_{- \lambda_{2}}(-p_z) ~ (RR_0^p)^{-1} \nonumber\\
&&    =\sum_{m_{S}^{'},\lambda_1,\lambda_2,\lambda} C_{s_1\lambda_{1},s_2\;-\lambda_{2}}^{S,\lambda}D^{S}_{m_{S}^{'},\lambda}(R R_0^p)~(R R_0^p)~H^{(1)}_{ \lambda_{1}}(p_z)H^{(2)}_{ -\lambda_{2}}(-p_z) ~ (RR_0^p)^{-1}.   \nonumber  
\eea  
The first part of (\ref{28}) contains the   spherical harmonics, which is expressed  in terms of the Wigner-D functions  as
\bea
\label{18}
&& \sum_{m_L,m_S} C^{Jm_J}_{Lm_L,Sm_S}Y^*_{Lm_L}(\widehat{Rp}) = \sum_{m_L,m_S} C^{Jm_J}_{Lm_L,Sm_S}Y^*_{Lm_L}(\widehat{R R_0^p p_z}) =  \\
&&= \sqrt{\frac{2L+1}{4 \pi}} \sum_{\lambda=-S}^{S}\sum_{m_S}  C^{J \lambda}_{L0,S \lambda} (D_{m_S, \lambda}^{S}(R R_0^p))^{*} D_{ m_J,\lambda }^{J}(R R_0^p).\nonumber 
\eea
which will be derived at the end of this Appendix. 

The partial wave operator now reads 
\bea
\label{first_op}
&&   O^{|p|,J,m_J,S,L}   = \sqrt{\frac{2L+1}{4 \pi}}\sum_{\lambda = -S}^{S} \sum_{\lambda_1,\lambda_2}  \sum_{R \in O^{(2)}}\sum_{m_S,m_{S}^{'}} C_{s_1\lambda_1,s_2\;-\lambda_2}^{S\lambda}C^{J \lambda}_{L0,S \lambda} \\
&&\qquad\qquad \times \ D^{S}_{m_{S}^{'},\lambda}(R R_0^p) (D_{m_S, \lambda}^{S}(R R_0^p))^{*} D_{ m_J,\lambda }^{J}(R R_0^p) \  (RR_0^p)~H^{(1)}_{ \lambda_{1}}(p_z)H^{(2)}_{- \lambda_{2}}(-p_z) ~(RR_0^p)^{-1}~ ,  \nonumber
\eea
which can be further simplified with use of Wigner-D function properties
\footnote{ $\sum_{\lambda} D^{S}_{m_{S}^{'},\lambda}(R R_0^p) (D_{m_S, \lambda}^{S}(R R_0^p))^{*}= \delta_{m_S,m_S^{'}}$ and    $
		D_{ m_J,\lambda }^{J}(R R_0^p)=\sum_{\lambda^{'}}D_{ m_J,\lambda^{'} }^{J}(R) D_{\lambda^{'},\lambda }^{J}(R_0^p)$.  } 
\bea
\label{con_without_P}
   O^{|p|,J,m_J,S,L}   &=& \sqrt{\frac{2L+1}{4 \pi}}\sum_{\lambda = -S}^{S} \sum_{\lambda_1,\lambda_2}C_{s_1\lambda_1,s_2\;-\lambda_2}^{S\lambda}C^{J \lambda}_{L0,S \lambda} \\
 \qquad \qquad &\times& \ \sum_{\lambda^{'}} D_{\lambda^{'},\lambda }^{J}(R_0^p) \sum_{R \in O^{(2)}} D_{ m_J,\lambda^{'} }^{J}(R) (RR_0^p)~H^{(1)}_{\lambda_{1}}(p_z)H^{(2)}_{ -\lambda_{2}}(-p_z) ~(RR_0^p)^{-1}~ . \nonumber
\eea
This operator already has a good parity $P=P_1P_2(-1)^L$ and one can act on it by  the parity projection $\frac{1}{2}(O+PIOI)$ without modifying  it,
\bea
&&  O^{|p|,J,m_J,S,L}   = \sqrt{\frac{2L+1}{4 \pi}}\sum_{\lambda = -S}^{S} \sum_{\lambda_1,\lambda_2}\sum_{\lambda^{'}} D_{\lambda^{'},\lambda }^{J}(R_0^p) ~C_{s_1\lambda_1,s_2-\lambda_2}^{S\lambda}~C^{J \lambda}_{L0,S \lambda}  \\
&&  \times ~\frac{1}{2} \sum_{R \in O^{(2)}} D_{ m_J,\lambda^{'} }^{J}(R) ~(RR_0^p) ~(H^{(1)}_{ \lambda_{1}}(p_z)H^{(2)}_{ -\lambda_{2}}(-p_z)+ PI H^{(1)}_{ \lambda_{1}}(p_z)H^{(2)}_{ -\lambda_{2}}(-p_z)I) ~(RR_0^p)^{-1}~. \nonumber 
\eea
The second line represents the helicity operator (\ref{O_helicity}), so the partial wave operator is the following linear combination of helicity operators:
\begin{equation}
\label{29}
 O^{|p|,J,m_J,S,L}  = \sqrt{\frac{2L+1}{4 \pi}} \sum_{\lambda = -S}^{S} \sum_{\lambda_{1},\lambda_{2}} \sum_{\lambda^{'}} D_{\lambda^{'},\lambda }^{J}(R_0^p)  C^{J\lambda}_{L0,S \lambda} C_{s_1\lambda_1,s_2-\lambda_2}^{S\lambda}~ O^{|p|,J,m_J,\lambda^{'},\lambda_1,\lambda_2}. ~~  
\end{equation}
A similar relation can be found in  Eq. (4) of \cite{Dudek:2012gj}. 

We close by deriving  the relation  (\ref{18}), which was used above.  The intermediate steps are taken from \cite{Varshalovich:1988}. Relation $Y^{LS}_{Jm_J}(\widehat{Rp})= \sum_{m_L,m_S} C^{Jm_J}_{Lm_L,Sm_S}Y_{Lm_L}(\widehat{Rp})\chi_{Sm_S}$ 
(page 196 of \cite{Varshalovich:1988}) 
 renders via $ \chi^{\dagger}_{S,m_S^\prime } \chi_{S,m_S}=\delta_{m_sm_s^\prime}$ the desired product  \begin{equation}
\label{36}
\sum_{m_L,m_S} C^{Jm_J}_{Lm_L,Sm_S}Y^*_{Lm_L}(\widehat{Rp})=\sum_{m_S}(\chi_{Sm_S}^{\dagger}Y^{LS}_{Jm_J}(\widehat{Rp}))^\dagger. 
\end{equation} 
The tensor $ Y^{LS}_{Jm_J}(\widehat{Rp})$ can be transformed  to helicity basis (denoted by prime on page 197 of \cite{Varshalovich:1988}) which in turn can be expressed using Wigner-D matrices (page 197 of \cite{Varshalovich:1988}) 
\begin{equation} 
	Y^{LS}_{Jm_J}(\theta,\phi)=\sum_{\lambda=-S}^{S}[Y_{Jm_J}^{LS}(\theta,\phi)]^{'\lambda}~ \chi_{S\lambda}^{h}(\theta,\phi) \nonumber
\end{equation}
where the right-hand-side can be further expressed as (pages 173 and 197 of \cite{Varshalovich:1988})
 \begin{equation} 
[Y_{Jm_J}^{LS}(\theta,\phi)]^{'\lambda} = \sqrt{\frac{2L+1}{4 \pi}} C^{J\lambda}_{L0,S\lambda}D_{-\lambda -m_J}^{J}(0,\theta,\phi) ,\quad \chi_{S\lambda}^{h}(\theta,\phi)=\sum_{m_S^{'}}D^{S}_{m_S^{'}\lambda}(\phi,\vartheta,0)\chi_{Sm_S^{'}}  
\end{equation}
Inserting all to (\ref{36}) and taking into account $D_{-\lambda -m_J}^{J}(\alpha,\beta,\gamma)=D_{m_J\lambda }^{J}(\gamma,\beta,\alpha)$ one arrives at 
\begin{equation}
\sum_{m_L,m_S} C^{Jm_J}_{Lm_L,Sm_S}Y^*_{Lm_L}(\theta, \phi)= \sqrt{\frac{2L+1}{4 \pi}} \sum_{\lambda=-S}^{S}\sum_{m_S}  C^{J \lambda}_{L0,S \lambda} (D_{m_S, \lambda}^{S}(\phi,\theta,0))^{*} D_{ m_J,\lambda }^{J}(\phi,\theta,0)~.\nonumber
\end{equation}
This is equivalent to (\ref{18}) for polar angles $\theta=\cos ^{-1}[(\widehat{Rp})_z/| \widehat{Rp}| ]$ and $\phi=-i \log [((\widehat{Rp})_x+i(\widehat{Rp})_y)/\sqrt{(\widehat{Rp})_x^2+(\widehat{Rp})_y^2}]$ of  $Rp=R R_0^p p_z$. The 
rotation of $p_z$ with Euler angles $(\phi,\theta,0)$  leads to the vector with polar coordinates $(\theta,\phi)$.

%\bibliography{bib/Lgt}
%\bibliographystyle{JHEP}

 \providecommand{\href}[2]{#2}\begingroup\raggedright\endgroup

\end{document}